\documentclass[lettersize,journal]{IEEEtran} \def\TwoCol

\IEEEoverridecommandlockouts

\usepackage{cite}
\usepackage{amsmath,amssymb,amsfonts}
\usepackage{tikz}
\usetikzlibrary{patterns}
\usetikzlibrary{decorations.pathreplacing}
\usetikzlibrary{shapes.geometric}
\usepackage{pgfplots}
\usetikzlibrary{calc}
\usetikzlibrary{arrows.meta}
\usepackage{amsmath}
\usepackage{amsfonts}
\usepackage{amssymb}
\usepackage{mathtools}
\usepackage{algorithmic}
\usepackage{graphicx}
\usepackage{textcomp}
\usepackage{blkarray}
\usepackage{array}
\usepackage{multirow}
\usepackage{arydshln}
\usepackage{bbm}
\usepackage{cite}
\usepackage{amsmath,amssymb,amsfonts}
\usepackage{tikz}
\usetikzlibrary{patterns}
\usetikzlibrary{decorations.pathreplacing}
\usetikzlibrary{matrix,decorations.pathreplacing, calc, positioning}
\usetikzlibrary{shapes.geometric}
\usepackage{pgfplots}
\usetikzlibrary{calc}
\usetikzlibrary{arrows.meta}
\usepackage{amsmath}
\usepackage{amsfonts}
\usepackage{arydshln}
\usepackage{amssymb}
\usepackage{mathtools}
\usepackage{algorithmic}
\usepackage{graphicx}
\usepackage{textcomp}
\usepackage{blkarray}
\usepackage{array}
\usepackage{multirow}
\usetikzlibrary{patterns}
\usepackage{mdframed}

\newcommand\numberthis{\addtocounter{equation}{1}\tag{\theequation}}
\usepackage{xcolor}
\def\BibTeX{{\rm B\kern-.05em{\sc i\kern-.025em b}\kern-.08em
    T\kern-.1667em\lower.7ex\hbox{E}\kern-.125emX}}
    
\makeatletter
\newcommand*{\rom}[1]{\expandafter\@slowromancap\romannumeral #1@}
\makeatother

\begin{document}

\title{\huge Age of Actuation and Timeliness: Semantics in a Wireless Power Transfer System}
\author{Ali Nikkhah, \IEEEmembership{Student Member, IEEE}, Anthony Ephremides, \IEEEmembership{Life Fellow, IEEE}, and Nikolaos Pappas, \IEEEmembership{Senior Member, IEEE}
\thanks{A. Nikkhah and N. Pappas are with the Department of Computer and Information Science Linköping University, Sweden, email: \{\texttt{ali.nikkhah, nikolaos.pappas\}@liu.se}. A. Ephremides is with the Electrical and Computer Engineering, University of Maryland, College Park, MD, USA, email: \texttt{etony@umd.edu}. \\ The work of A. Nikkhah and N. Pappas has been supported in part by the Swedish Research Council (VR), ELLIIT, Zenith, and the European Union (ETHER, 101096526). A shorter version has been published in \cite{nikkhah2023age}.}
}

\maketitle
\begin{abstract}
In this paper, we investigate a model relevant to semantics-aware goal-oriented communications, and we propose a new metric that incorporates the utilization of information in addition to its timelines. Specifically, we consider the transmission of observations from an external process to a battery-powered receiver through status updates. These updates inform the receiver about the process status and enable actuation if sufficient energy is available to achieve a goal. We focus on a wireless power transfer (WPT) model, where the receiver receives energy from a dedicated power transmitter and occasionally from the data transmitter when they share a common channel. We analyze the Age of Information (AoI) and propose a new metric, the \textit{Age of Actuation (AoA), which is relevant when the receiver utilizes the status updates to perform actions in a timely manner}. We provide analytical characterizations of the average AoA and the violation probability of the AoA, demonstrating that AoA generalizes AoI. Moreover, we introduce and analytically characterize the \textit{Probability of Missing Actuation (PoMA)}; this metric becomes relevant also \textit{to quantify the incurred cost of a missed action}. We formulate unconstrained and constrained optimization problems for all the metrics and present numerical evaluations of our analytical results. This proposed set of metrics goes beyond the traditional timeliness metrics since the synergy of different flows is now considered.
\end{abstract}

\section{Introduction}

In the current communication systems, the importance and timing of information are often neglected; thus, generated information often becomes irrelevant or not important for the task at hand. Consequently, redundant information is transmitted over the network, increasing the unnecessary usage of network and energy resources. Semantics of information (SoI) and goal-oriented communications is an emerging research direction that takes into account the importance and relevance of information to achieve a specific goal \cite{kountouris2021semantics, pappas2021goal, gunduz2022beyond, Timing6G,utkovski2023semantics} by considering the entire information chain from data generation to transmission and utilization. This new paradigm is relevant in the emerging wireless ecosystem where a large number of interacting devices are connected, such as in Wireless Network Control Systems (WNCS) \cite{WNCSSurvey}, and not all information packets are equally important. The ultimate aim of SoI is to reduce or even eliminate the generation and transmission of unnecessary or stale data packets in communication systems, thereby mitigating redundant network congestion, resource exploitation, and excessive power consumption \cite{kountouris2021semantics, kutsevol2023experimental}.

In \cite{kountouris2021semantics, pappas2021goal}, it was proposed that information attributes can be categorized as innate and contextual. A considered innate metric in semantics-aware goal-oriented communications, the age of information (AoI) \cite{Kosta2017age, TutorialYates, pappas2023age} played a pivotal role in motivating this new paradigm. AoI measures the freshness of received data packets at a destination, capturing the timeliness of information in status-updating systems. However, the AoI solely focuses on the freshness of information and does not directly address the timeliness of actuation based on the received data. This paper introduces a metric called the \textit{Age of Actuation (AoA)}, which measures the elapsed time since the last actuation at a remote destination based on received data. To enable actuation, the destination requires wireless energy from a power-transmitting device. Additionally, we propose the \textit{Probability of Missing Actuation (PoMA)} as another metric, which penalizes missed actions where the actuator fails to respond to the transmitted data; PoMA can also capture the cost of a missed action. By introducing these metrics, we aim to address the timeliness of actuation in goal-oriented communications. Note that the AoA metric introduced in this paper differs from the query age of information \cite{chiariotti2022query}, where the age is considered only when needed and queried. In contrast, AoA is continuously considered and resets only when energy and data are provided.

Radio frequency wireless power transfer (WPT) \cite{varshney2008transporting} plays a crucial role in networks where energy harvesting devices cannot be connected to the power grid and rely on remote energy sources. It is particularly suitable for long-distance energy transfer \cite{lu2015wireless}, making it an ideal solution for remote applications. In recent years, the significance of addressing both the timeliness of information and energy harvesting (EH) has driven numerous studies in energy harvesting systems. However, most of these studies have focused on energy consumption while transmitting information rather than leveraging the information to perform actions.

\subsection{Related Work}
We focus on the characterization and modeling of energy harvesting (EH) systems, focusing on their involvement in models related to our topic. For a comprehensive discussion on the AoI and its applications, we refer the reader to \cite{pappas2023age}. It is important to note that in almost all related studies, the harvested energy is predominantly consumed for transmitting data packets. EH systems can be broadly classified into two types. The first type models EH as an external stochastic process separate from the environment. In continuous-time models, EH is often modeled using a Poisson process \cite{wu2017optimal, farazi2018age, YatesISIT2015, feng2018optimal, zheng2019closed, gindullina2021age, bacingolu2019optimal, ArafaTWC2019, elmagid2022age, elmagid2022closed, arafa2019age}, while discrete-time models employ a Bernoulli process \cite{pappas2020average,dong2020energy, hatami2021aoi, chen2021optimization, CeranAoI2019, xie2022age, XuTMC23}.

The second type of EH involves harvesting energy from WPT systems. Usually, these models consider energy accumulation required for transmission during the time. One notable approach in such systems is adopting greedy or best-effort policies, where the transmission is triggered once the accumulated energy reaches a certain threshold \cite{krikidis2019average}. Alternatively, energy accumulation without a greedy policy is also explored \cite{ibrahim2016stability}, and energy can be assumed to arrive in discrete units \cite{moradian2020stability}. Markov Decision Process (MDP) formulations are also employed to address the problem \cite{abd2020aoi, sheikhi2021aoi}. In a different study on energy accumulation and greedy policies \cite{leng2019ageof}, the energy accumulation process is categorized into three states. Motivated by the non-linear characteristics of EH circuits such as sensitivity and saturation \cite{boshkovska2015practical}, energy levels below a threshold are considered insufficient for charging, levels above a threshold are saturated and capped at a constant value, and intermediate levels behave linearly. However, the probability distribution for these three intervals is determined by the geometrical adjacency of the parties rather than the randomness of fading power. It is worth noting that while the aforementioned papers focus on using harvested energy to transmit information, they do not specifically address how to utilize the received information to perform an action, which is the focus of this study.

\subsection{Contributions}

Our model focuses on a battery-powered receiver that relies on received status updates to gain insights into the status of an external process and execute actuation. While receiving status updates does not consume energy, the actuation process does. The model incorporates two transmission nodes: a dedicated power-transmitting node that wirelessly provides energy to the receiver and a data-transmitting node responsible for status updates. Both nodes operate on the same wireless channel, leading to power and data transmission interference. This configuration falls under simultaneous wireless information and power transfer (SWIPT) technology, which benefits interference for efficient wireless power transfer \cite{zhang2013mimo, ibrahim2016stability}. In particular, we adopt a co-located SWIPT setup, where a single antenna serves energy and data reception.

The two primary architectures for SWIPT are time switching (TS) and power splitting (PS) \cite{zhang2013mimo}. In the TS architecture, the transmission switches between data and energy, while in the PS architecture, the receiver performs simultaneous decoding of information and energy harvesting by splitting the received power into two fractions \cite{park2017analysis, shi2014joint, tuan2019optimizing}. Our model incorporates aspects from both architectures. Successful data transmission occurs when the signal-to-interference-and-noise ratio (SINR) exceeds a predefined threshold. Similarly, for energy transmission, we consider success when the received energy at the destination is above a certain threshold. Our work shares similarities with \cite{leng2019ageof}, but with the distinction that our model considers two intervals below sensitivity and above saturation. Moreover, we accurately calculate the probabilities of successful energy packet charging based on the stochastic nature of fading channels.

Below we summarize our contributions.
\begin{itemize}
\item We propose a model relevant to semantics-aware goal-oriented communications, where status updates generated by a device inform a remote destination about the status of an external process. The battery-powered destination relies on these updates to perform specific actions and achieve its goals. Our model incorporates SWIPT, where the destination receives energy from a dedicated power transmitter and the data transmitter.

\item We analyze the AoI at the receiver and we provide an analytical characterization of the battery's evolution, which allows us to define and analyze the AoA metric. We demonstrate that AoA is a more general metric than AoI, particularly relevant when the receiver utilizes status updates to perform actions in a timely manner, which is crucial in goal-oriented communications. We also analyze the violation probability of both the AoI and the AoA and introduce the metric of the PoMA to penalize unexecuted/missed actions.

\item We address the optimization of several performance metrics, including the average AoI, average AoA, violation probabilities of AoI and AoA, and the constrained optimization of PoMA with an average AoI threshold. Furthermore, we explore the optimization of the average AoI while imposing an average AoA threshold. To complement our theoretical analysis, we provide numerical evaluations of our analytical results, offering insights into the performance and efficiency of our proposed metrics and optimization approaches.
\end{itemize}

\section{System Model} \label{System Model}
We consider a system model consisting of two transmitting devices as depicted in Fig. \ref{SystemModelFig}; the first transmits status updates as data packets to a receiver, and the second is dedicated to power transmission. The devices share the same wireless channel under random access. The receiver performs an actuation upon reception of a data packet, given that there is available stored energy in its battery in terms of energy packets. The transmitters provide the receiver with the necessary data and energy for performing an actuation. Thus, we have SWIPT, which consists of energy harvesting (EH) and information decoding (ID) units at the receiver side. The two transmitters and the receiver are equipped with a single antenna. The parameters $\rho$ and $\rho^2$, as illustrated in Fig. \ref{SystemModelFig}, denote the signal and power splitting ratio, respectively. Specifically, $\rho^2$ and $1-\rho^2$ are the split and directed power fractions to the EH and ID circuits. Time is discrete and divided into timeslots, $t$, of the same length. The first transmitter observes a random process and samples and transmits a status update with probability (w.p.) $q_1$ in a timeslot. This status update will inform the receiver about the status of the remote source. Then, it will be utilized to perform an action if there is sufficient energy.
The receiver can harvest energy from both transmitters, as explained below. The second transmitter transmits wireless power w.p. $q_2$ in a time slot. While transmitting data, the first transmitter can also assist with the energy when only the second transmitter transmits. In the case of a successful transmission of energy, about which more details will be given later, an energy packet is stored in the battery. We consider the cases of finite-sized and infinite-sized batteries.

At the end of a time slot, the receiver can perform an actuation by utilizing one energy packet from its battery and using the received data packet in the same time slot. However, the receiver cannot perform an action without data reception or insufficient energy. Thus, an actuation is performed in a time slot according to the following two cases: i) if there is a non-empty battery and there is a successful data reception, ii) there is an empty battery, but there are successful receptions of both energy and data packets during that time slot.

\ifdefined\TwoCol
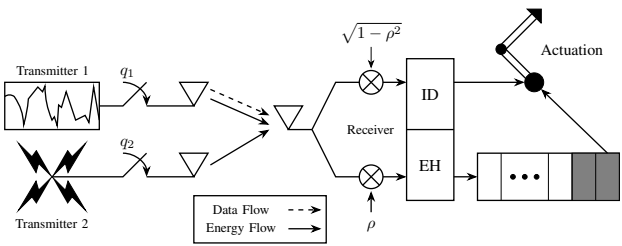
\begin{figure}[b]
\centering
\resizebox{0.46\textwidth}{!}{
\begin{tikzpicture}

\draw [thick] (0,0) rectangle (2,1) node[scale=0.8] [xshift=-1.23cm,yshift=0.3cm] {Transmitter 1}; 
\draw [thick] (0,0.7) to [bend right=-35] (0.2,0.7) to [bend right=-15]  (0.4,0.1) to [bend right=5] (0.5,0.65) -- (0.7,0.9) -- (0.8,0.8) -- (0.85,0.92) to [bend right=10] (1,0.2)-- (1.05,0.27) -- (1.11,0.08) --(1.14,0.52) -- (1.2,0.47) -- (1.3,0.8) -- (1.5,0.2) -- (1.7,0.3) -- (1.78,0.15) -- (1.88,0.9) -- (2,0.2);

\draw [thick] (2,0.5) -- (2.5,0.5) -- (3,1);
\draw[thick] (3,0.5) -- (4,0.5) -- (3.7,1) -- (4.3,1) -- (4,0.5);
\draw [-Stealth,thick] (2.5,1) to [bend right=-45] (3,0.5) node [xshift=-0.4cm, yshift=0.7cm] {$q_1$};
\draw [-Stealth, thick, dashed] (4.35,0.85) -- (5.6,0.3);
\draw [-Stealth, thick] (4.35,0.65) -- (5.6,0.1);

\draw[thick] (4,-2.4) rectangle (6.8,-1.4) node[xshift=-1.8cm,yshift=-0.3cm, scale=0.8] {Data Flow} node[xshift=-1.8cm,yshift=-0.7cm, scale=0.8] {Energy Flow} ;
\draw[-Stealth,thick, dashed] (6,-1.7) -- (6.7,-1.7);
\draw[-Stealth,thick] (6,-2.1) -- (6.7,-2.1);

\node[scale=0.8] [xshift=1.25cm,yshift=-2.5cm] {Transmitter 2};

\draw[fill=black, thin, scale=1, xshift=1cm , yshift=-1cm] (0,0) -- (45:0.4cm) -- (15:0.35cm) -- (45:1cm) -- (45:0.6cm) -- (60:0.65);

\begin{scope}[yscale=1,xscale=-1]
\draw[fill=black, thin, scale=1, xshift=-1cm , yshift=-1cm] (0,0) -- (45:0.4cm) -- (15:0.35cm) -- (45:1cm) -- (45:0.6cm) -- (60:0.65);
\end{scope}

\begin{scope}[yscale=-1,xscale=1]
\draw[fill=black, thin, scale=1, xshift=1cm , yshift=1cm] (0,0) -- (45:0.4cm) -- (15:0.35cm) -- (45:1cm) -- (45:0.6cm) -- (60:0.65);
\end{scope}

\begin{scope}[yscale=-1,xscale=-1]
\draw[fill=black, thin, scale=1, xshift=-1cm , yshift=1cm] (0,0) -- (45:0.4cm) -- (15:0.35cm) -- (45:1cm) -- (45:0.6cm) -- (60:0.65);
\end{scope}

\draw[thick] (1,-1) -- (2.5,-1) -- (3,-0.5);
\draw[thick] (3,-1) -- (4,-1) -- (3.7,-0.5) -- (4.3,-0.5) -- (4,-1);
\draw [-Stealth, thick] (2.5,-0.5) to [bend right=-45] (3,-1) node [xshift=-0.4cm, yshift=0.7cm] {$q_2$};
\draw [-Stealth, thick] (4.35,-0.75) -- (5.6,0);

\draw[thick] (6,0) -- (5.7,0.5) -- (6.3,0.5) -- (6,0) -- (6.5,0);

\draw[thick] (6.5,0) -- (7,1) -- (7.5,1);
\draw[thick] (7.75,1) circle (0.25cm);
\draw[thick] (7.75,1) -- (7.93,1.18);
\draw[thick] (7.75,1) -- (7.93,0.82);
\draw[thick] (7.75,1) -- (7.57,1.18);
\draw[thick] (7.75,1) -- (7.57,0.82);

\draw[-Stealth] (7.75,1.75) -- (7.75,1.3) node [xshift=0cm, yshift=0.75cm] {$\sqrt{1-\rho^2}$};
\draw[-Stealth, thick] (8,1) -- (8.5,1);

\begin{scope}[yscale=-1,xscale=1]
\draw[thick] (6.5,0) -- (7,1) -- (7.5,1);
\draw[thick] (7.75,1) circle (0.25cm);
\draw[thick] (7.75,1) -- (7.93,1.18);
\draw[thick] (7.75,1) -- (7.93,0.82);
\draw[thick] (7.75,1) -- (7.57,1.18);
\draw[thick] (7.75,1) -- (7.57,0.82);

\draw[-Stealth, thick] (7.75,1.75) -- (7.75,1.3) node [xshift=0cm, yshift=-0.75cm] {$\rho$};
\draw[-Stealth, thick] (8,1) -- (8.5,1);
\end{scope}

\draw [thick] (8.5,1.5) rectangle (9.5,-1.5);
\draw [thick] (8.5,0) -- (9.5,0) node [xshift=-0.5cm, yshift=0.75cm] {ID} node [xshift=-0.5cm, yshift=-0.75cm] {EH} node[scale=0.8] [xshift=-2.2cm, yshift=0cm] {Receiver};

\draw[-Stealth, thick] (9.5,1) -- (11,1);
\draw[fill,black] (11.2,1) circle (0.2cm);
\draw[thick] (11.25,1.05) -- (11.25-0.7,1.05+0.7);
\draw[thick] (11.15,0.95) -- (11.15-0.7,0.95+0.7);
\draw[fill,black] (10.50,1.70) circle (0.11cm);
\node [ xshift=12cm, yshift=1.70cm,scale=0.95] {Actuation};
\draw[thick] (10.55,1.65) -- (10.55+0.7,1.65+0.7);
\draw[thick] (10.45,1.75) -- (10.45+0.7,1.75+0.7);

\draw[fill=black] (10.55+0.8,1.65+0.6) -- (10.45+0.6,1.75+0.8) -- (10.55+0.8,1.75+0.8)-- (10.55+0.8,1.65+0.6);

\draw[-Stealth, thick] (9.5,-1) -- (10,-1);
\draw[thick] (10,-0.5) rectangle (13,-1.5);
\draw[fill=gray] (12,-0.5) rectangle (13,-1.5);
\draw (12.5,-1.5) -- (12.5,-0.5);
\draw (12,-1.5) -- (12,-0.5);
\draw (11.5,-1.5) -- (11.5,-0.5);
\draw (10.5,-1.5) -- (10.5,-0.5);

\draw[fill,black] (10.75,-1) circle (0.05cm);
\draw[fill,black] (11,-1) circle (0.05cm);
\draw[fill,black] (11.25,-1) circle (0.05cm);

\draw[-Stealth, thick] (12.75,-0.5) -- (11.34,0.86);

\end{tikzpicture}
}
\caption{The considered system model.} 
\label{SystemModelFig}
\end{figure}
\fi

\ifdefined\OneCol
\begin{figure}[b]
\centering
\resizebox{0.8\textwidth}{!}{
\begin{tikzpicture}

\draw [thick] (0,0) rectangle (2,1) node[scale=0.8] [xshift=-1.23cm,yshift=0.3cm] {Transmitter 1}; 
\draw [thick] (0,0.7) to [bend right=-35] (0.2,0.7) to [bend right=-15]  (0.4,0.1) to [bend right=5] (0.5,0.65) -- (0.7,0.9) -- (0.8,0.8) -- (0.85,0.92) to [bend right=10] (1,0.2)-- (1.05,0.27) -- (1.11,0.08) --(1.14,0.52) -- (1.2,0.47) -- (1.3,0.8) -- (1.5,0.2) -- (1.7,0.3) -- (1.78,0.15) -- (1.88,0.9) -- (2,0.2);

\draw [thick] (2,0.5) -- (2.5,0.5) -- (3,1);
\draw[thick] (3,0.5) -- (4,0.5) -- (3.7,1) -- (4.3,1) -- (4,0.5);
\draw [-Stealth,thick] (2.5,1) to [bend right=-45] (3,0.5) node [xshift=-0.4cm, yshift=0.7cm] {$q_1$};
\draw [-Stealth, thick, dashed] (4.35,0.85) -- (5.6,0.3);
\draw [-Stealth, thick] (4.35,0.65) -- (5.6,0.1);

\draw[thick] (4,-2.4) rectangle (6.8,-1.4) node[xshift=-1.8cm,yshift=-0.3cm, scale=0.8] {Data Flow} node[xshift=-1.8cm,yshift=-0.7cm, scale=0.8] {Energy Flow} ;
\draw[-Stealth,thick, dashed] (6,-1.7) -- (6.7,-1.7);
\draw[-Stealth,thick] (6,-2.1) -- (6.7,-2.1);

\node[scale=0.8] [xshift=1.25cm,yshift=-2.5cm] {Transmitter 2};

\draw[fill=black, thin, scale=1, xshift=1cm , yshift=-1cm] (0,0) -- (45:0.4cm) -- (15:0.35cm) -- (45:1cm) -- (45:0.6cm) -- (60:0.65);

\begin{scope}[yscale=1,xscale=-1]
\draw[fill=black, thin, scale=1, xshift=-1cm , yshift=-1cm] (0,0) -- (45:0.4cm) -- (15:0.35cm) -- (45:1cm) -- (45:0.6cm) -- (60:0.65);
\end{scope}

\begin{scope}[yscale=-1,xscale=1]
\draw[fill=black, thin, scale=1, xshift=1cm , yshift=1cm] (0,0) -- (45:0.4cm) -- (15:0.35cm) -- (45:1cm) -- (45:0.6cm) -- (60:0.65);
\end{scope}

\begin{scope}[yscale=-1,xscale=-1]
\draw[fill=black, thin, scale=1, xshift=-1cm , yshift=1cm] (0,0) -- (45:0.4cm) -- (15:0.35cm) -- (45:1cm) -- (45:0.6cm) -- (60:0.65);
\end{scope}

\draw[thick] (1,-1) -- (2.5,-1) -- (3,-0.5);
\draw[thick] (3,-1) -- (4,-1) -- (3.7,-0.5) -- (4.3,-0.5) -- (4,-1);
\draw [-Stealth, thick] (2.5,-0.5) to [bend right=-45] (3,-1) node [xshift=-0.4cm, yshift=0.7cm] {$q_2$};
\draw [-Stealth, thick] (4.35,-0.75) -- (5.6,0);

\draw[thick] (6,0) -- (5.7,0.5) -- (6.3,0.5) -- (6,0) -- (6.5,0);

\draw[thick] (6.5,0) -- (7,1) -- (7.5,1);
\draw[thick] (7.75,1) circle (0.25cm);
\draw[thick] (7.75,1) -- (7.93,1.18);
\draw[thick] (7.75,1) -- (7.93,0.82);
\draw[thick] (7.75,1) -- (7.57,1.18);
\draw[thick] (7.75,1) -- (7.57,0.82);

\draw[-Stealth] (7.75,1.75) -- (7.75,1.3) node [xshift=0cm, yshift=0.75cm] {$\sqrt{1-\rho^2}$};
\draw[-Stealth, thick] (8,1) -- (8.5,1);

\begin{scope}[yscale=-1,xscale=1]
\draw[thick] (6.5,0) -- (7,1) -- (7.5,1);
\draw[thick] (7.75,1) circle (0.25cm);
\draw[thick] (7.75,1) -- (7.93,1.18);
\draw[thick] (7.75,1) -- (7.93,0.82);
\draw[thick] (7.75,1) -- (7.57,1.18);
\draw[thick] (7.75,1) -- (7.57,0.82);

\draw[-Stealth, thick] (7.75,1.75) -- (7.75,1.3) node [xshift=0cm, yshift=-0.75cm] {$\rho$};
\draw[-Stealth, thick] (8,1) -- (8.5,1);
\end{scope}

\draw [thick] (8.5,1.5) rectangle (9.5,-1.5);
\draw [thick] (8.5,0) -- (9.5,0) node [xshift=-0.5cm, yshift=0.75cm] {ID} node [xshift=-0.5cm, yshift=-0.75cm] {EH} node[scale=0.8] [xshift=-2.2cm, yshift=0cm] {Receiver};

\draw[-Stealth, thick] (9.5,1) -- (11,1);
\draw[fill,black] (11.2,1) circle (0.2cm);
\draw[thick] (11.25,1.05) -- (11.25-0.7,1.05+0.7);
\draw[thick] (11.15,0.95) -- (11.15-0.7,0.95+0.7);
\draw[fill,black] (10.50,1.70) circle (0.11cm);
\node [ xshift=12cm, yshift=1.70cm,scale=0.95] {Actuation};
\draw[thick] (10.55,1.65) -- (10.55+0.7,1.65+0.7);
\draw[thick] (10.45,1.75) -- (10.45+0.7,1.75+0.7);

\draw[fill=black] (10.55+0.8,1.65+0.6) -- (10.45+0.6,1.75+0.8) -- (10.55+0.8,1.75+0.8)-- (10.55+0.8,1.65+0.6);

\draw[-Stealth, thick] (9.5,-1) -- (10,-1);
\draw[thick] (10,-0.5) rectangle (13,-1.5);
\draw[fill=gray] (12,-0.5) rectangle (13,-1.5);
\draw (12.5,-1.5) -- (12.5,-0.5);
\draw (12,-1.5) -- (12,-0.5);
\draw (11.5,-1.5) -- (11.5,-0.5);
\draw (10.5,-1.5) -- (10.5,-0.5);

\draw[fill,black] (10.75,-1) circle (0.05cm);
\draw[fill,black] (11,-1) circle (0.05cm);
\draw[fill,black] (11.25,-1) circle (0.05cm);

\draw[-Stealth, thick] (12.75,-0.5) -- (11.34,0.86);

\end{tikzpicture}
}
\caption{The considered system model.} 
\label{SystemModelFig}
\end{figure}
\fi

We denote $\cal{D}$ the event of a successful data packet transmission in a time slot that occurs w.p.
\begin{equation} \label{P_u1}
P_{\cal{D}}=q_1q_2P_{d12}+q_1\bar{q}_2P_{d1},
\end{equation}
where $P_{d1}$ is the success probability for the data packets when only the first transmitter attempts to transmit; $P_{d12}$ is the success probability when both transmitters transmit simultaneously. Also, note that $\bar{q}_{i}=1-q_{i}$. The expressions are given in Section \ref{The Success Probabilities of Transmissions in Different Scenarios}. We assume that if a packet is not successfully transmitted in a slot, then it is dropped by the transmitter. In this study, we do not require the presence of a feedback channel from the receiver to the transmitter.

We denote $\cal{E}$ the event of a successful transmission of an energy packet in a time slot, and it occurs w.p.
\begin{equation} \label{P_u2}
P_{\cal{E}}=q_1q_2P_{e12}+\bar{q}_1q_2P_{e2},
\end{equation}
where $P_{e12}$, $P_{e2}$ are the probabilities of a successful transmission of an energy packet when both of the transmitters transmit and when only the second transmitter transmits, respectively. 
The expressions are given in Section \ref{The Success Probabilities of Transmissions in Different Scenarios}.

The probabilities of four possible joint outcomes are
\begin{align}
 \label{P_u1_u2}
&P_{\cal{D,E}}=q_1 q_2 P_{d12} P_{e12},\\ 
 \label{P_u1_u2b}
&P_{\cal{D,\bar{E}}}= q_1 q_2 P_{d12} \bar{P}_{e21} +q_1 \bar{q}_2 P_{d1},\\
\label{P_u1b_u2}
&P_{\cal{\bar{D},E}}=q_1 q_2 \bar{P}_{d12} P_{e12} + \bar{q}_1 q_2 P_{e2},\\
\label{P_u1b_u2b}
&P_{\cal{\bar{D},\bar{E}}}=\bar{q}_1\bar{q}_2+q_1 \bar{q}_2 \bar{P}_{d1} + \bar{q}_1 q_2 \bar{P}_{e2} + q_1 q_2 \bar{P}_{d12} \bar{P}_{e12}.
\end{align}
Note that $P_{\cal{D}}=P_{\cal{D,E}}+P_{\cal{D,\bar{E}}}$, $P_{\cal{E}}=P_{\cal{D,E}}+P_{\cal{\bar{D},E}}$, and $P_{\cal{D,E}}+P_{\cal{D,\bar{E}}}+P_{\cal{\bar{D},E}}+P_{\cal{\bar{D},\bar{E}}}=1$.

\subsection{Physical Layer Model} \label{Physical Layer Model}
The \textit{time switching} technique is \textit{switching} between the two extremes of the PS technique, i.e., $\rho=0$ and $\rho=1$, along the \textit{time} slots \cite{shi2014joint}. Hence, we consider three cases $\rho=0$, $\rho=1$, and $0 < \rho <1$, altering from time slot to time slot, depending on the possible events of sole or co-transmission of the transmitters as discussed in section \ref{The Success Probabilities of Transmissions in Different Scenarios}. A general SWIPT model can be found in many papers in the literature, such as \cite{park2017analysis,shi2014joint,tuan2019optimizing}.
In a SWIPT system mentioned above, the $SINR$ of the $j$-th transmitter at the receiver, $j \in \{1,2\}$, is given by \cite{park2017analysis}
\begin{equation} \label{SNIR_PS}
SINR=\frac{ d_{i}^{-\alpha_i} |h_{i}^{*} w_i |^2 }{\sum_{j=1, j\neq i}^{2} d_{j}^{-\alpha_j} |h_{j}^{*} w_j |^2 + \frac{\sigma_n^2}{K_s P_s} }.
\end{equation}
in which $K_s=(1-\rho^2)$ and $K_s=1$ under the PS and TS methods, respectively.
The harvested energy, using a linear model for EH and considering unit size time slot, at the receiver, is \cite{park2017analysis}
\begin{equation} \label{HE_PS}
E= K_h P_s \left( \sum_{j=1}^{2}  d_j^{-\alpha_j} |h_{j}^{*} w_j |^2 \right),
\end{equation}
where $K_h=\rho^2$ and $K_h=1$ for the PS and the TS method, respectively. Note that the radio frequency to direct circuit conversion efficiency is assumed to be $1$.
The $P_s=E[|s_j^2|]=1$ is assumed \cite{tuan2019optimizing}. We denote the noise power with $\sigma_n^2=P_N$. There is the term
\ifdefined\OneCol
\begin{equation}
|h_{j}^{*} w_j |^2=|h_{j}^{*}|^2 |w_j |^2=|h_{j}|^2 |w_j |^2 ,
\end{equation}
\fi
\ifdefined\TwoCol
\begin{align}
|h_{ji}^{*} w_j |^2&=|h_{ji}^{*}|^2 |w_j |^2 ,\\
&=|h_{ji}|^2 |w_j |^2 ,
\end{align}
\fi
in which $|w_j|^2=||w_j||_2^2$ is the power of the $j$th transmitter. Since $h_{ji}$ is a normal complex random variable (RV) with zero mean and the variance of $\sqrt{(\frac{\upsilon}{2})}^2$, its absolute value, i.e., $R=|h_{ji}|$, is a Rayleigh RV with the probability density function (PDF) $f_R(x)=\frac{2x}{\upsilon}\text{exp}(-\frac{x^2}{\upsilon})$ \cite{papoulis2002probability}. Also, $A=|h_{ji}|^2$ has the PDF of $f_A (x) = \frac{1}{\upsilon} \text{exp} (-\frac{x}{\upsilon}), x \geq 0$ \cite{nguyen2008optimization}. Hence, we have $ |\mathbf{h}_{ji}^H \mathbf{w}_j |^2= A_j P_{tx,j}$, in which $A_j \sim A(\upsilon)$ denotes the small-scale fading for the link of the $j$-th transmitter, and in this work, we assumed that is Rayleigh fading. $P_{tx,j}$ is the transmission power of the $j$th transmitter. Then, $P_{rx,j}=d_j^{-\alpha} A_j P_{tx,j}= g_j A_j$, where $g_j=P_{tx,j} d_j^{-\alpha}$ is the power factor, and $P_{rx,j}$ is the received power from the $j$-th transmitter. The $d_j$ is the distance between the $j$-th transmitter and the receiver, and $\alpha_j$ is the path loss exponent from the $j$-th transmitter to the receiver.
\subsection{Success Probabilities of Data and Energy Transmissions} \label{The Success Probabilities of Transmissions in Different Scenarios}
We consider a successful transmission of a data packet when the signal-to-interference-and-noise ratio (SINR) of the data transmission is above a threshold, i.e., $SINR \geq \gamma_d$. We also consider a successful transmission of an energy packet whenever the harvested energy at the receiver is above a threshold, i.e., $E \geq \gamma_e$.
The probabilities of successful transmissions of data and energy packets are given below.
\subsubsection{Both Transmitters Transmit ($0 \leq \rho \leq 1$)} \label{Both Transmitters Transmit}
A data packet transmission is successful when 
\begin{equation}
SINR_{d12}=\frac{P_{rx,1}}{P_{rx,2}+\frac{P_N}{1-\rho^2}} \geq \gamma_d
\end{equation}
and it occurs w.p.
\begin{equation} \label{T_d12}
\displaystyle{P_{d12}=\text{exp}\left(-\frac{\gamma_d  P_N}{(1-\rho^2) \upsilon_1  g_1}\right) \left(1+\gamma_d \frac{\upsilon_2 g_2}{\upsilon_1 g_1}\right)^{-1}}.
\end{equation} 
It can be obtained by \cite[Theorem 1]{nguyen2008optimization}, dividing the noise power by $1-\rho^2$.
A successful transmission of an energy packet occurs when $E_{e12}= \rho^2(P_{rx,1} +  P_{rx,2}) \geq \gamma_e $, w.p.
\ifdefined\TwoCol
{\small
\begin{align*}
&P_{e12}=\\
&1-\frac{g_2 \upsilon_2 \left(\text{exp}\left(-\frac{\gamma_e}{\rho^2 g_2 \upsilon_2}\right)-1\right)+g_1 \upsilon_1 \left(1-\text{exp}\left(-\frac{\gamma_e}{\rho^2 g_1 \upsilon_1}\right)\right)}{g_1\upsilon_1 - g_2 \upsilon_2}. \numberthis \label{T_e12}
\end{align*}
}
\fi
\ifdefined\OneCol
\begin{equation} \label{T_e12}
P_{e12}=1-\frac{g_2 \upsilon_2 \left(\text{exp}\left(-\frac{\gamma_e}{\rho^2 g_2 \upsilon_2}\right)-1\right)+g_1 \upsilon_1 \left(1-\text{exp}\left(-\frac{\gamma_e}{\rho^2 g_1 \upsilon_1}\right)\right)}{g_1\upsilon_1 - g_2 \upsilon_2}.
\end{equation}
\fi
The proof is given in Appendix A.

\subsubsection{Only the First Transmitter Transmits ($ \rho = 0$)} \label{Only the First Transmitter Transmits}
In this case, if $SINR_{d1}=\frac{P_{rx,1}}{P_N} \geq \gamma_d$, there is a successful data transmission \cite[Theorem 1]{nguyen2008optimization}, w.p. $P_{d1}=\text{exp}\left(-\frac{\gamma_d P_N}{g_1 \upsilon_1  } \right)$.

\subsubsection{Only the Second Transmitter Transmits ($\rho=1$)} \label{Only the Second Transmitter Transmits}
A successful energy transmission occurs when $E_{e2}=P_{rx,2} \geq \gamma_e$ and its probability is given by $P_{e2}= \text{exp} \left(- \frac{\gamma_e}{ g_2 \upsilon_2}\right)$
which is obtained from (\ref{T_e12}), after setting $g_1=0$ and $\rho=1$.

\section{Age of Information (AoI)} 

This section presents the mathematical analysis to characterize AoI and the AoI violation probability under the generate-at-will policy for the considered system model. Furthermore, we consider a set of optimization problems for the mentioned metrics.

\subsection{Analysis} \label{The Evolution of Age of Information}

The reception of data packets is independent of energy availability, as mentioned in section \ref{System Model}. Thus, at time slot $t+1$, the value of the AoI denoted by $I(t+1)$ drops to $1$ if there is a successful data reception. Otherwise, AoI, $I(t+1)$, increases by $1$. A sample path of AoI is depicted by vertical lines in Fig. \ref{AoA_AoI_Evo}. The AoI evolution is summarized by
\begin{equation} \label{AoI_evo}
I(t+1)=\begin{cases}
1 & \text{w.p.} \  P_{\cal{D}} , \\
I(t)+1 & \text{w.p.} \  1-P_{\cal{D}} . \\
\end{cases}
\end{equation}
AoI can be modeled as a Discrete Time Markov Chain (DTMC)\cite{fountoulakis2022information}, with a Markov transition probability matrix 
\begin{equation} \label{TPM_AoI}
\mathbf{P}^I=
\begin{bmatrix}
P_{\cal{D}} & 1- P_{\cal{D}} &0 & 0 &0\\
P_{\cal{D}} & 0 & 1-P_{\cal{D}} & 0& 0\\
P_{\cal{D}} &  0& 0 & 1-P_{\cal{D}} & 0\\
\vdots & \vdots & \vdots & \vdots & \vdots\\ 
\end{bmatrix} .
\end{equation}

To obtain the steady state of this Markov process, we solve the system of $\mathbf{\pi}^I \mathbf{P}^I =\mathbf{\pi}^I$ and $\mathbf{\pi}^I \mathbf{1}=1$. The first equation is $P_{\cal{D}}(\pi_1^I+\pi_2^I+\hdots)=\pi_1^I$ which leads to $\pi_1^I=P_{\cal{D}}$. Other equations would be $(1-P_{\cal{D}})\pi_i^I=\pi_{i+1}^I$ for $i\geq 1$. Recursively, it can be seen that $\pi_i^I=P_{\cal{D}}(1-P_{\cal{D}})^{i-1}$.
The term $\pi_i^I$ is the probability of the AoI having the value $i$ at each time slot. Then, the average AoI is
\begin{align*} 
\bar{I}&=\sum_{i=1}^{\infty}\pi_i^I i=\sum_{i=1}^{\infty}P_{\cal{D}}(1-P_{\cal{D}})^{i-1}i=\frac{P_{\cal{D}}}{1-P_{\cal{D}}}\sum_{i=1}^{\infty}(1-P_{\cal{D}})^i i\\
&\overset{(a)}{=}\frac{1}{P_{\cal{D}}}, \numberthis \label{AoI_mean}
\end{align*}
in which (a) is due to $\sum_{i=1}^{\infty}ic^i=\frac{c}{(1-c)^2}, \forall |c|<1$.

We can also define the AoI violation probability as the probability that the AoI is greater than a threshold, e.g., $I_v$, and is given by
\begin{align*} 
V_I&=\sum_{i=I_v+1}^{\infty} \pi_{i}^{I} = 1- \sum_{i=1}^{I_v} \pi_{i}^{I}=1-\sum_{i=1}^{I_v} P_{\cal{D}} (1-P_{\cal{D}})^{i-1}\\
&=1-P_{\cal{D}}(1-(1-P_{\cal{D}})^{I_v}). \numberthis \label{viol_aoi}
\end{align*}

\ifdefined\TwoCol
\begin{figure*}
\hrule
\vspace{1em}
\begin{equation} \label{TPM_B}
\mathbf{P}=
\begin{bmatrix}
P_{\cal{D,E}}+P_{\cal{D,\bar{E}}}+P_{\cal{\bar{D},\bar{E}}} & P_{\cal{\bar{D},E}} & 0 & 0 & \hdots & 0\\
P_{\cal{D,\bar{E}}} & P_{\cal{D,E}}+P_{\cal{\bar{D},\bar{E}}} & P_{\cal{\bar{D},E}} &0 & 
\hdots & 0 \\
0 & P_{\cal{D,\bar{E}}} & P_{\cal{D,E}}+P_{\cal{\bar{D},\bar{E}}} & P_{\cal{\bar{D},E}} & \hdots & 0 \\
\vdots & \ddots & \ddots & \ddots & \ddots & \vdots \\ \cdashline{1-6} 
0& \hdots & 0 & P_{\cal{D,\bar{E}}} & P_{\cal{D,E}}+P_{\cal{\bar{D},\bar{E}}} & P_{\cal{\bar{D},E}} \\
0& \hdots & 0 & 0 & P_{\cal{D,\bar{E}}} & P_{\cal{D,E}}+P_{\cal{\bar{D},E}}+P_{\cal{\bar{D},\bar{E}}}\\
\end{bmatrix} 
\end{equation}
\hrule
\end{figure*}
\fi
\ifdefined\OneCol
\begin{equation} \label{TPM_B}
\begin{aligned}
\mathbf{P}&=
\left[\begin{matrix}
P_{\cal{D,E}}+P_{\cal{D,\bar{E}}}+P_{\cal{\bar{D},\bar{E}}} & P_{\cal{\bar{D},E}} & 0 \\

P_{\cal{D,\bar{E}}} & P_{\cal{D,E}}+P_{\cal{\bar{D},\bar{E}}} & P_{\cal{\bar{D},E}} \\

0 & P_{\cal{D,\bar{E}}} & P_{\cal{D,E}}+P_{\cal{\bar{D},\bar{E}}} \\

\vdots & \ddots & \ddots  \\ \cdashline{1-6}

0 & 0 & 0 \\

0& 0 & 0  \\

\end{matrix}\right.\\
&\qquad\qquad
\left.\begin{matrix}

 0 & \hdots & 0\\

0 &  \hdots & 0 \\

P_{\cal{\bar{D},E}} & \hdots & 0 \\

\ddots & \ddots & \vdots\\ \cdashline{1-6} 

P_{\cal{D,\bar{E}}} & P_{\cal{D,E}}+P_{\cal{\bar{D},\bar{E}}} & P_{\cal{\bar{D},E}} \\

0 &   P_{\cal{D,\bar{E}}} & P_{\cal{D,E}}+P_{\cal{\bar{D},E}}+P_{\cal{\bar{D},\bar{E}}}\\

\end{matrix}\right].
\end{aligned}
\end{equation}
\fi

\subsection{Optimization of the Average AoI} \label{Age of Information_opt}
We optimize the average AoI with the objective function $\bar{I}\left(q_1,q_2\right)=\frac{1} {{\mathcal{P}}_{D}}=\bar{A}_1$. The gradient vector with respect to $q_1$ and $q_2$, is
\ifdefined\TwoCol
\begin{align*} 
\nabla\bar{I}\left(q_1,q_2\right)=\Bigg[ &-\frac{P_{d1} \left(1 - q_2\right) + P_{d12} q_2}{\left(P_{d1} q_1 \left(1 - q_2\right) + P_{d12} q_1 q_2\right)^2}  , \\ 
& \frac{ q_1\left(P_{d1}-P_{d12}\right) }{\left(P_{d1} q_1 \left(1 - q_2\right) + P_{d12} q_1 q_2\right)^2} \Bigg]. \numberthis \label{Grad_Inf_AoI}
\end{align*}
\fi
\ifdefined\OneCol
\begin{equation} \label{Grad_Inf_AoI}
\nabla\bar{I}\left(q_1,q_2\right)=\left[-\frac{P_{d1} \left(1 - q_2\right) + 
P_{d12} q_2}{\left(P_{d1} q_1 \left(1 - q_2\right) + P_{d12} q_1 q_2\right)^2},\frac{ q_1\left(P_{d1}-P_{d12}\right) }{\left(P_{d1} q_1 \left(1 - q_2\right) + P_{d12} q_1 q_2\right)^2}\right].
\end{equation} 
\fi

Given that $0 \leq q_1, q_2 \leq 1$, the first element is always negative. Also, since $P_{d1} \geq P_{d12}$, the second element is always positive. Thus, the minimum average AoI is achieved by values
\begin{equation} \label{Opt_points_AoI}
[q_1^{*},q_2^{*}]=[1,0],   
\end{equation}
and the optimum value is
\begin{equation} \label{Opt_value_AoI}
\bar{I}^{*}= \frac{1}{P_{d1}}.
\end{equation}

\textit{Remark 1:} Since AoI does not depend on energy, to minimize the average AoI and the violation probability of AoI, we need to silence the transmitter dedicated to the power transmission to eliminate the interference. However, this will not be the case in a constrained optimization problem of minimizing the average AoI under the average AoA constraint. 

For the optimization of the AoI violation probability, refer to Remark 3 in section \ref{Optimization of the Age of Actuation}.

\section{Age of Actuation (AoA)}

In this section, we first present the analysis of the evolution of the battery and then define the metric of the AoA, which becomes relevant when status updates are used to perform energy-required actions in a timely manner. With this metric, we depart from the classical freshness metrics and go towards \textit{goal-oriented semantics-aware communication systems} where we may need more than one packet to act \textit{synergistically} to perform an action. One packet consists of the required information and the other with the required energy to perform the action based on the data. The AoA captures the time since the last actuation was performed. We also characterize the violation probability of AoA. We conclude the section by considering two optimization problems.

\subsection{The Evolution of the Battery} \label{The Evolution of the Battery}

We consider two cases, the finite-sized and the infinite-sized batteries. We denote the state of the battery at a time slot $t$ by $B(t)$. Then we have $B(t)=\{0,1,\hdots,m\}$ and $B(t)=\{0,1,\hdots,\infty\}$ for the two cases of finite-sized ($m$ is the size of the battery) and infinite-sized batteries, respectively. The evolution of the battery is described below.
If $B(t)=0$, then
\begin{equation} \label{B_evo_first}
B(t+1)=\begin{cases}
0 & \text{w.p.} \ P_{\cal{D,E}}+ P_{\cal{D,\bar{E}}}+P_{\cal{\bar{D},\bar{E}}} , \\
1 & \text{w.p.} \  P_{\cal{\bar{D},E}} . \\
\end{cases}
\end{equation}
In the finite case if $ 1 \leq B(t) \leq m-1$, and in the infinite case if $B(t) \geq 1$, we have
\begin{equation} \label{B_evo_middle}
B(t+1)=\begin{cases}
B(t)-1 & \text{w.p.} \  P_{\cal{D,\bar{E}}} , \\
B(t) & \text{w.p.} \  P_{\cal{D,E}}+P_{\cal{\bar{D},\bar{E}}} , \\
B(t)+1 & \text{w.p.} \  P_{\cal{\bar{D},E}} . \\
\end{cases}
\end{equation}
If $B(t)=m$, in the finite-sized battery we have
\begin{equation}  \label{B_evo_last}
B(t+1)=\begin{cases}
m-1 & \text{w.p.} \  P_{\cal{D,\bar{E}}} , \\
m & \text{w.p.} \  P_{\cal{D,E}}+P_{\cal{\bar{D},E}}+P_{\cal{\bar{D},\bar{E}}} . \\
\end{cases}
\end{equation}

A DTMC can model the battery size evolution for both cases; its transition probability matrix is given by (\ref{TPM_B}).

The whole matrix is considered for the finite case, whereas the matrix above the dashed line can represent the infinite case. The steady-state distribution is represented by $\boldsymbol{\pi}$, and $\pi_k$ is the probability that the battery has stored $k$ energy packets. The probability that the battery is empty, $\mathrm{Pr}(B=0)$, is denoted by $\pi_0$ and $\bar{\pi}_0$ denotes the probability of a non-empty battery, $\mathrm{Pr}(B\neq 0)$. To simplify the notation we denote the event of a non-empty battery by ${\mathcal{B}}$. The expressions for the probabilities mentioned above are provided in the following.

The infinite-sized and finite-sized batteries can be modeled by a Geo/Geo/1 queue and a Geo/Geo/1/B queue, respectively \cite[Sections 3.4.2 and 3.4.3]{srikant2013communication}. In \cite{srikant2013communication}, service utilization is defined as the probability of one arrival and no departure divided by the probability of one departure and no arrival. One can see that the former probability, in the battery, is $P_{\cal{\bar{D},E}}$ and the latter probability is $P_{\cal{D,\bar{E}}}$. Hence, for the infinite case, the probability of an empty battery is $\mathrm{Pr}(B=0)=1-\frac{P_{\bar{{\cal{D}}},{\cal{E}}}}{P_{{\cal{D}},\bar{{\cal{E}}}}}$ and the probability of a non-empty battery is $\mathrm{Pr}(B\neq 0)=\frac{P_{\bar{{\cal{D}}},{\cal{E}}}}{P_{{\cal{D}},\bar{{\cal{E}}}}}$, respectively.
Note that, for the infinite-sized battery, these are valid when the stability condition, i.e., $P_{\bar{{\cal{D}}},{\cal{E}}} < P_{{\cal{D}},\bar{{\cal{E}}}}$, holds. When $P_{\bar{{\cal{D}}},{\cal{E}}} \geq P_{{\cal{D}},\bar{{\cal{E}}}}$, then the battery never empties which is equivalent to a system that is always connected to the power grid. Thus we have
\begin{equation} \label{empty_p_inf}
\pi_0^i=
\begin{cases}
0 & \frac{P_{\bar{{\cal{D}}},{\cal{E}}}}{P_{{\cal{D}},\bar{{\cal{E}}}}} \geq 1, \\
1- \frac{P_{\bar{{\cal{D}}},{\cal{E}}}}{P_{{\cal{D}},\bar{{\cal{E}}}}}  & \frac{P_{\bar{{\cal{D}}},{\cal{E}}}}{P_{{\cal{D}},\bar{{\cal{E}}}}} < 1, \\
\end{cases}
\quad
\bar{\pi }_0^i=
\begin{cases}
1 & \frac{P_{\bar{{\cal{D}}},{\cal{E}}}}{P_{{\cal{D}},\bar{{\cal{E}}}}} \geq 1, \\
\frac{P_{\bar{{\cal{D}}},{\cal{E}}}}{P_{{\cal{D}},\bar{{\cal{E}}}}} & \frac{P_{\bar{{\cal{D}}},{\cal{E}}}}{P_{{\cal{D}},\bar{{\cal{E}}}}} < 1. \\
\end{cases}
\end{equation}

For the finite-sized battery, the probabilities of empty and non-empty batteries are
\begin{equation} \label{empty_p_fin}
\pi_0^f=\frac{1-\frac{P_{\bar{{\cal{D}}},{\cal{E}}}}{P_{{\cal{D}},\bar{{\cal{E}}}}}}{1-\left(\frac{P_{\bar{{\cal{D}}},{\cal{E}}}}{P_{{\cal{D}},\bar{{\cal{E}}}}}\right)^{m+1}} , 
\quad
\bar{\pi}_0^f=\frac{\frac{P_{\bar{{\cal{D}}},{\cal{E}}}}{P_{{\cal{D}},\bar{{\cal{E}}}}}-\left(\frac{P_{\bar{{\cal{D}}},{\cal{E}}}}{P_{{\cal{D}},\bar{{\cal{E}}}}}\right)^{m+1}}{1-\left(\frac{P_{\bar{{\cal{D}}},{\cal{E}}}}{P_{{\cal{D}},\bar{{\cal{E}}}}}\right)^{m+1}} .  
\end{equation}
Also, the probability of the finite-sized battery being full is
\begin{equation} \label{full_p_fin}
\pi_m^f=\frac{\left(\frac{P_{\bar{{\cal{D}}},{\cal{E}}}}{P_{{\cal{D}},\bar{{\cal{E}}}}}\right)^m-\left(\frac{P_{\bar{{\cal{D}}},{\cal{E}}}}{P_{{\cal{D}},\bar{{\cal{E}}}}}\right)^{m+1}}{1-\left(\frac{P_{\bar{{\cal{D}}},{\cal{E}}}}{P_{{\cal{D}},\bar{{\cal{E}}}}}\right)^{m+1}} . 
\end{equation}
Note that, the condition of $\frac{P_{\bar{{\cal{D}}},{\cal{E}}}}{P_{{\cal{D}},\bar{{\cal{E}}}}} < 1$ is required only for the infinite size case. 
We can see that (\ref{empty_p_fin}) transform to (\ref{empty_p_inf}) as $m \rightarrow \infty$, when $\frac{P_{\bar{{\cal{D}}},{\cal{E}}}}{P_{{\cal{D}},\bar{{\cal{E}}}}} < 1$. We use the superscripts $i$ and $f$ to distinguish between \textit{infinite}-sized and \textit{finite}-sized batteries, respectively. 
\subsection{The Analysis of AoA} \label{The Evolution of Actuation Freshness}

\begin{figure}
\centering 
\scalebox{.9}{ \boldmath{
\begin{tikzpicture}

\fill[pattern=horizontal lines] (0,0) rectangle (1,1);
\fill[pattern=horizontal lines] (1,0) rectangle (2,2);
\fill[pattern=horizontal lines] (2,0) rectangle (3,3);
\fill[pattern=horizontal lines] (3,0) rectangle (4,4);
\fill[pattern=horizontal lines] (4,0) rectangle (5,5);
\fill[pattern=horizontal lines] (5,0) rectangle (6,1);
\fill[pattern=horizontal lines] (6,0) rectangle (7,2);

\fill[pattern=vertical lines] (0,0) rectangle (1,1);
\fill[pattern=vertical lines] (1,0) rectangle (2,2);
\fill[pattern=vertical lines] (2,0) rectangle (3,1);
\fill[pattern=vertical lines] (3,0) rectangle (4,2);
\fill[pattern=vertical lines] (4,0) rectangle (5,3);
\fill[pattern=vertical lines] (5,0) rectangle (6,1);
\fill[pattern=vertical lines] (6,0) rectangle (7,2);

\draw[-Stealth, very thick ] (0,0) -- (8,0)  node [xshift=0cm, yshift=-0.5cm] {$t$} node [xshift=-8cm, yshift=-0.5cm] {$0$} node [xshift=-6cm, yshift=-0.7cm] {$({\mathcal{D}},\bar{{\mathcal{E}}} \cap \bar{{\mathcal{B}}} )$} node [xshift=-3cm, yshift=-0.7cm] {$({\mathcal{D}},{\mathcal{E}} \cup {\mathcal{B}} )$};

\draw [decorate,decoration={brace,amplitude=5pt,raise=0.5ex}, thick] (1,-0.5) --  (3,-0.5);
\draw [decorate,decoration={brace,amplitude=5pt,raise=0.5ex}, thick] (4,-0.5) --  (6,-0.5);

\draw[-Stealth, very thick ] (0,0) -- (0,6) node [xshift=-0.7cm, yshift=0cm] {$A(t)$} node [xshift=-0.7cm, yshift=-0.6cm] {$I(t)$} node [xshift=-0.4cm, yshift=-5cm] {$1$} node [xshift=-0.4cm, yshift=-6cm] {$0$};

\draw (1,-0.1) -- (1,0.1);
\draw (2,-0.1) -- (2,0.1);
\draw (3,-0.1) -- (3,0.1);
\draw (4,-0.1) -- (4,0.1);
\draw (5,-0.1) -- (5,0.1);
\draw (6,-0.1) -- (6,0.1);
\draw (7,-0.1) -- (7,0.1);

\draw (-0.1,1) -- (0.1,1);
\draw (-0.1,2) -- (0.1,2);
\draw (-0.1,3) -- (0.1,3);
\draw (-0.1,4) -- (0.1,4);
\draw (-0.1,5) -- (0.1,5);
\draw (6,-0.1) -- (6,0.1);

\draw[very thick] (0,1) -- (1,1);

\draw[very thick] (1,1) -- (1,2);
\draw[very thick] (1,2) -- (2,2);

\draw[very thick] (2,2) -- (2,3);
\draw[very thick] (2,3) -- (3,3);

\draw[very thick] (3,3) -- (3,4);
\draw[very thick] (3,4) -- (4,4);

\draw[very thick] (4,4) -- (4,5);
\draw[very thick] (4,5) -- (5,5);

\draw[ very thick] (5,5) -- (5,1);
\draw[very thick] (5,1) -- (6,1);

\draw[ very thick] (6,1) -- (6,2);
\draw[ very thick] (6,2) -- (7,2);

\draw[dashed, very thick] (2,2) -- (2,1);
\draw[dashed, very thick] (2,1) -- (3,1);

\draw[dashed, very thick] (3,1) -- (3,2);
\draw[dashed, very thick] (3,2) -- (4,2);

\draw[dashed, very  thick] (4,2) -- (4,3);
\draw[dashed, very thick] (4,3) -- (5,3);

\node[xshift=6.2cm,yshift=5cm] {AoA};
\node[xshift=6.2cm,yshift=4.5cm] {AoI};

\fill[pattern=horizontal lines] (6.7,4.8) rectangle (7.7,5.2);
\fill[pattern=vertical lines] (6.7,4.3) rectangle (7.7,4.7);

\draw (5.7,4.2) rectangle (7.9,5.3);

\end{tikzpicture}}} 
\caption{The evolution of AoA and AoI metrics.}
\label{AoA_AoI_Evo}
\end{figure}
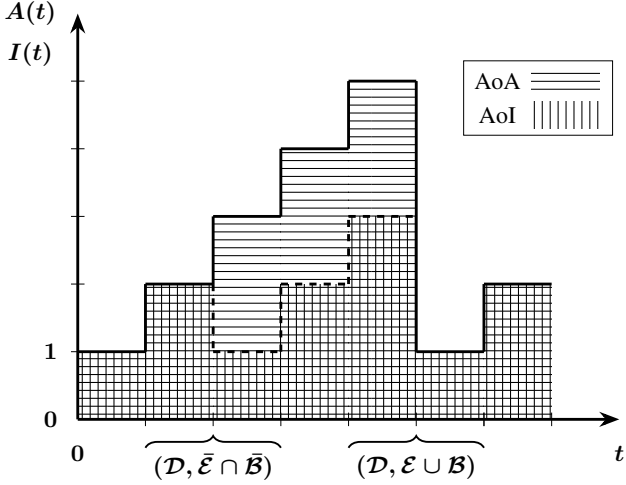

AoA, $A(t)$, is the time elapsed since the last actuation was performed; thus, $A(t)=t-a(t)$, where $a(t)$ denotes the time of the last performed actuation. We assume that $A(0)=1$. In Fig. \ref{AoA_AoI_Evo}, we depict a sample path of AoA. The area of the vertical lines represents the AoI, and the area of the horizontal lines depicts the AoA. Thus, the grid area is the overlap. A data packet is successfully transmitted at $t=2$ and $t=5$, and the AoI is reset to $1$. However, at $t=2$, the energy transmission is unsuccessful, and the battery is empty. Thus, the AoA continues to grow. At $t=5$, either the energy transmission is successful or the battery is non-empty. Hence, AoA is reset to $1$. In other time slots, there is no successful transmission of a data packet, i.e., $\bar{{\mathcal{D}}}$. 

In a time slot $t+1$, $A(t+1)=1$ when in time $t$ has happened one of the following two cases. First, when there is a successful transmission of data, i.e., ${\cal{D}}$, and also the battery is non-empty, i.e., ${\mathcal{B}}$, thus it can provide the energy for an actuation. It occurs w.p. $P_{\cal{D}} \bar{\pi}_0^b$. Second, if the battery is empty, i.e., $\bar{{\mathcal{B}}}$, then we need the joint event $({\cal{D}},{\cal{E}})$, meaning that we have simultaneously successful energy and data transmissions. It occurs w.p. $P_{{\cal{D}},{\cal{E}}} \pi_0^b$. Other than these two cases, $A(t+1)=A(t)+1$. The evolution of the AoA can be summarized by
\begin{equation} \label{AF_evo}
A(t+1)=\begin{cases}
1 & \text{w.p.} \  P_{\cal{D}} \bar{\pi}_0^b + P_{\cal{D},\cal{E}} \pi_0^b , \\
A(t)+1 & \text{w.p.} \  1-\left(P_{\cal{D}} \bar{\pi}_0^b + P_{\cal{D},\cal{E}} \pi_0^b\right) . \\
\end{cases}
\end{equation}
Note that this evolution is similar to (\ref{AoI_evo}) except it is $P_{\cal{D}} \bar{\pi}_0^b + P_{\cal{D},\cal{E}} \pi_0^b$ instead of $P_{\cal{D}}$. Hence, from (\ref{AoI_mean}), we have
\begin{equation} \label{AF_mean_2}
\bar{A}=\frac{1}{P_{\cal{D}} \bar{\pi}_0^b + P_{\cal{D},\cal{E}} \pi_0^b}.
\end{equation}
For the infinite-sized battery, by replacing $\pi_0^b$ and $\bar{\pi}_0^b$ from (\ref{empty_p_inf}), we obtain
\begin{equation} \label{AF_b_inf}
\bar{A}=
\begin{cases}
\bar{A}_1=\frac{1}{P_{\cal{D}}} & \frac{P_{\bar{{\cal{D}}},{\cal{E}}}}{P_{{\cal{D}},\bar{{\cal{E}}}}} \geq 1,\\
\bar{A}_2=\frac{1}{ P_{\cal{D}} \frac{P_{\bar{{\cal{D}}},{\cal{E}}}}{P_{{\cal{D}},\bar{{\cal{E}}}}} + P_{{\cal{D}},{\cal{E}}} \left(1-\frac{P_{\bar{{\cal{D}}},{\cal{E}}}}{P_{{\cal{D}},\bar{{\cal{E}}}}}\right)  }
 & \frac{P_{\bar{{\cal{D}}},{\cal{E}}}}{P_{{\cal{D}},\bar{{\cal{E}}}}} < 1.\\
\end{cases}
\end{equation}
For the finite-sized battery, by replacing $\pi_0^b$ and $\bar{\pi}_0^b$ from (\ref{empty_p_fin}), we obtain
\begin{equation} \label{AF_b_fin}
\bar{A}=\frac{1-\left(\frac{P_{\bar{{\cal{D}}},{\cal{E}}}}{P_{{\cal{D}},\bar{{\cal{E}}}}}\right)^{m+1}}{P_{\cal{D}} \left(\frac{P_{\bar{{\cal{D}}},{\cal{E}}}}{P_{{\cal{D}},\bar{{\cal{E}}}}}-\left(\frac{P_{\bar{{\cal{D}}},{\cal{E}}}}{P_{{\cal{D}},\bar{{\cal{E}}}}}\right)^{m+1}\right)+ P_{{\cal{D}},{\cal{E}}}\left(1-\frac{P_{\bar{{\cal{D}}},{\cal{E}}}}{P_{{\cal{D}},\bar{{\cal{E}}}}}\right) } . 
\end{equation}
Note that (\ref{AF_b_fin}) transforms to (\ref{AF_b_inf}) as $m \rightarrow \infty$, when $\frac{P_{\bar{{\cal{D}}},{\cal{E}}}}{P_{{\cal{D}},\bar{{\cal{E}}}}} < 1$.

We can also obtain the violation probability of AoA. Considering (\ref{viol_aoi}), and replacing $P_{\cal{D}}$ by $P_{\cal{D}} \bar{\pi}_0^b + P_{\cal{D},\cal{E}} \pi_0^b$, we have
{\small
\begin{equation} \label{viol_aoa}
V_A=1-\left(P_{\cal{D}} \bar{\pi}_0^b + P_{\cal{D},\cal{E}} \pi_0^b\right)\left(1+\left(1-\left(P_{\cal{D}} \bar{\pi}_0^b + P_{\cal{D},\cal{E}} \pi_0^b\right)\right)^{I_v}\right),
\end{equation}
}
in which $A_v$ is the violation limit of the AoA.
For infinite-sized batteries, from (\ref{empty_p_inf}) and (\ref{viol_aoa}), we have

{\scriptsize
\begin{equation} \label{viol_aoa_infi}
V_A=
\begin{cases}
1-P_{\cal{D}}+P_{\cal{D}}\left(1-P_{\cal{D}}\right)^{I_v} & \frac{P_{\bar{{\cal{D}}},{\cal{E}}}}{P_{{\cal{D}},\bar{{\cal{E}}}}} \geq 1,\\
1-\left(P_{\cal{D}} \frac{P_{\bar{{\cal{D}}},{\cal{E}}}}{P_{{\cal{D}},\bar{{\cal{E}}}}} + P_{\cal{D},\cal{E}} \left(1-\frac{P_{\bar{{\cal{D}}},{\cal{E}}}}{P_{{\cal{D}},\bar{{\cal{E}}}}}\right)\right)\times \\
\left(1+\left(1-\left(P_{\cal{D}} \frac{P_{\bar{{\cal{D}}},{\cal{E}}}}{P_{{\cal{D}},\bar{{\cal{E}}}}} + P_{\cal{D},\cal{E}} \left(1-\frac{P_{\bar{{\cal{D}}},{\cal{E}}}}{P_{{\cal{D}},\bar{{\cal{E}}}}}\right)\right)\right)^{I_v}\right) & \frac{P_{\bar{{\cal{D}}},{\cal{E}}}}{P_{{\cal{D}},\bar{{\cal{E}}}}} < 1. \\
\end{cases}
\end{equation}
}

Then for finite-sized batteries, from (\ref{empty_p_fin}) and (\ref{viol_aoa}), we obtain
{\tiny
\begin{align*} 
V_A=&1-\left(P_{\cal{D}} \frac{\frac{P_{\bar{{\cal{D}}},{\cal{E}}}}{P_{{\cal{D}},\bar{{\cal{E}}}}}-\left(\frac{P_{\bar{{\cal{D}}},{\cal{E}}}}{P_{{\cal{D}},\bar{{\cal{E}}}}}\right)^{m+1}}{1-\left(\frac{P_{\bar{{\cal{D}}},{\cal{E}}}}{P_{{\cal{D}},\bar{{\cal{E}}}}}\right)^{m+1}} + P_{\cal{D},\cal{E}} \frac{1-\frac{P_{\bar{{\cal{D}}},{\cal{E}}}}{P_{{\cal{D}},\bar{{\cal{E}}}}}}{1-\left(\frac{P_{\bar{{\cal{D}}},{\cal{E}}}}{P_{{\cal{D}},\bar{{\cal{E}}}}}\right)^{m+1}}\right) \times \\
&\left(1+\left(1-\left(P_{\cal{D}} \frac{\frac{P_{\bar{{\cal{D}}},{\cal{E}}}}{P_{{\cal{D}},\bar{{\cal{E}}}}}-\left(\frac{P_{\bar{{\cal{D}}},{\cal{E}}}}{P_{{\cal{D}},\bar{{\cal{E}}}}}\right)^{m+1}}{1-\left(\frac{P_{\bar{{\cal{D}}},{\cal{E}}}}{P_{{\cal{D}},\bar{{\cal{E}}}}}\right)^{m+1}} + P_{\cal{D},\cal{E}} \frac{1-\frac{P_{\bar{{\cal{D}}},{\cal{E}}}}{P_{{\cal{D}},\bar{{\cal{E}}}}}}{1-\left(\frac{P_{\bar{{\cal{D}}},{\cal{E}}}}{P_{{\cal{D}},\bar{{\cal{E}}}}}\right)^{m+1}}\right)\right)^{I_v}\right). \numberthis \label{viol_aoa_fi}
\end{align*}
}

\textit{Remark 2:} When the system is not limited by energy, the AoA reduces to the AoI, as Fig. \ref{AoA_AoI_Evo} depicts. This is because energy will always be available to perform a potential action. Thus, AoA depends only on the arrival of data, as does AoI. Hence, this metric is a more general case of AoI, and it captures a semantic attribute relevant in goal-oriented semantics-aware communications when an actuation is involved.

\subsection{Optimization of the Average AoA} \label{Optimization of the Age of Actuation}
We consider the infinite case for the battery. Hence, the objective function is (\ref{AF_b_inf}).
For the scenario where the system is not energy limited, i.e., the first case in (\ref{AF_b_inf}), since $\bar{A}_1 = \bar{I}$, the analysis in Section \ref{Age of Information_opt} holds. Here we consider the $\bar{A}_2$. 
We have that

\ifdefined\TwoCol
\begin{align*}
\nabla\bar{A}_2 \left(q_1,q_2\right)=\Bigg[&\frac{P_{e2} - 
  P_{e12}}{\left(\left(q_{1}-1\right) P_{e2}  -  q_{1} P_{e12}\right)^{2} q_{2} },\\
  &\frac{1}{\left(\left(q_{1}-1\right)P_{e2}  -q_{1}P_{e12} \right) q_{2}^{2}}\Bigg]. \numberthis  \label{Grad_Inf_AoA}
\end{align*}
\fi
\ifdefined\OneCol
\begin{equation}  \label{Grad_Inf_AoA}
\nabla\bar{A}_2 \left(q_1,q_2\right)=[\frac{P_{e2} - 
P_{e12}}{\left(\left(q_{1}-1\right) P_{e2}  -  q_{1} P_{e12}\right)^{2} q_{2} },\frac{1}{\left(\left(q_{1}-1\right)P_{e2}  -q_{1}P_{e12} \right) q_{2}^{2}}].
\end{equation} 
\fi

The sign of the first element depends on $\left(P_{e2}-P_{e12}\right)$, and the second element is always negative.
The intersection of the two areas of $\bar{A}_1$ and $\bar{A}_2$ requires to solve the line $P_{\bar{{\cal{D}}},{\cal{E}}} =P_{{\cal{D}},\bar{{\cal{E}}}}$.
Then we obtain
\begin{equation} \label{Equ_Border} 
q_1  P_{d1} - q_2 P_{e2}   =  q_1 q_2( P_{d1} - P_{d12} + P_{e12} - P_{e2}). 
\end{equation}
We solve it for $q_1$ (or $q_2$) and then insert it into either $\bar{A}_1$ or $\bar{A}_2$ to obtain the curve of intersection. The plausible critical point of this curve is
\begin{equation} \label{theta1}
\theta_1=\frac{P_{e2}(P_{e2}  - P_{e12}) - \sqrt{(P_{d1} - P_{d12}) P_{e2}^2 (P_{e2} - P_{e12})}}{(P_{e12} - P_{e2}) (P_{d1} - P_{d12} + P_{e12} - P_{e2})},
\end{equation}
and
{\small
\begin{equation} \label{theta2}
\theta_2=\frac{P_{d1} (P_{e12} P_{e2} - P_{e2}^2 + \sqrt{(P_{d1} - P_{d12}) P_{e2}^2 (P_{e2} - P_{e12})})}{(P_{d1} - P_{d12} + P_{e12} -  P_{e2}) \sqrt{(P_{d1} - P_{d12}) (P_{e2} - P_{e12}) P_{e2}^2}} .
\end{equation}
}
Also, the intersection of the two areas intersects with the border of $q_1=1$ at
\begin{equation} \label{delta2}
\delta_2   =  \frac{P_{d1} }{  P_{d1} +P_{e12}- P_{d12} },
\end{equation} 
and intersects with the border of $q_2=1$ at
\begin{equation} \label{delta1}
\delta_1   =  \frac{P_{e2} }{  P_{e2} +P_{d12}- P_{e12} }.
\end{equation}

Comparing (\ref{Grad_Inf_AoI}) and (\ref{Grad_Inf_AoA}), if $P_{e12}>P_{e2}$, then the first elements of the gradients of $\bar{A}_1$ and $\bar{A}_2$ have the same sign, thus, the value of $\bar{A}$ would monotonously decrease with increasing of $q_1$. It can be confirmed by noting that (\ref{theta1}) and (\ref{theta2}) would be complex when $P_{e12}>P_{e2}$. Hence, $q_1^*=1$. Then, $q_2$ would be calculated by (\ref{delta2}). If $P_{e12} \geq P_{d12}$, then $\delta_2 \geq 1$ and hence $[q_1^*,q_2^*]=[1,1]$. If $P_{e12} < P_{d12}$, then $[q_1^*,q_2^*]=[1,\delta_2]$.

If $P_{e12}<P_{e2}$, then (\ref{theta1}) and (\ref{theta2}) would have rational values. If $\theta_1 \leq 1$ and $ \theta_2 \leq 1$, we have $[q_1^*,q_2^*]=[\theta_1,\theta_2]$. Otherwise, the minimum would be on the border of $q_1=1$ or $q_2=1$. From (\ref{delta2}) and (\ref{delta1}), it is observable that if $P_{e12} < P_{d12}$ then $[q_1^*,q_2^*]=[\delta_1,1]$, if $P_{e12} > P_{d12}$ then $[q_1^*,q_2^*]=[1,\delta_2]$, and if $P_{e12} = P_{d12}$ then $[q_1^*,q_2^*]=[1,1]$.

Thus, the minimum average AoA is summarized as
\begin{equation} \label{Opt_point_Ave_AoA}
[q_1^*,q_2^*]=
\begin{cases}
[1,1] & P_{e12} > P_{e2}, P_{d12} \geq P_{e12}, \\
[1,\delta_2] & P_{e12} > P_{e2}, P_{d12} < P_{e12}, \\
[\theta_1,\theta_2] & P_{e12} < P_{e2}, \{\theta_1 <1, \theta_2<1\},\\
[\delta_1,1] & P_{e12} < P_{e2}, \{\theta_1 <1, \theta_2>1\} \cup\\
&\{\theta_1 >1, \theta_2>1\},P_{d12} > P_{e12},\\
[1,\delta_2] & P_{e12} < P_{e2}, \{\theta_1 >1, \theta_2<1\} \cup\\
&\{\theta_1 >1, \theta_2>1\},P_{d12} < P_{e12},\\
[1,1] & P_{e12} < P_{e12} < P_{e2}, \{\theta_1 >1, \theta_2>1\},\\
&P_{d12} = P_{e12}.
\end{cases}
\end{equation}

The minimum average AoA after replacing (\ref{Opt_point_Ave_AoA}) in (\ref{AF_mean_2}) is 

\begin{equation} \label{Opt_value_Ave_AoA}
\bar{A}^*=
\begin{cases}
\frac{1}{P_{e12}} & P_{e12} > P_{e2}, P_{d12} \geq P_{e12}, \\
\frac{P_{d1} - P_{d12} + P_{e12}}{P_{d1} P_{e12}} & P_{e12} > P_{e2}, P_{d12} < P_{e12}, \\
\bar{I}(\theta_1,\theta_2) & P_{e12} < P_{e2}, \{\theta_1 <1, \theta_2<1\},\\
\frac{P_{d12} - P_{e12} + P_{e2}}{P_{d12} P_{e2}} & P_{e12} < P_{e2}, \{\theta_1 <1, \theta_2>1\} \cup\\
&\{\theta_1 >1, \theta_2>1\}, P_{d12} > P_{e12},\\
\frac{P_{d1} - P_{d12} + P_{e12}}{P_{d1} P_{e12}} & P_{e12} < P_{e2}, \{\theta_1 >1, \theta_2<1\} \cup\\
&\{\theta_1 >1, \theta_2>1\}, P_{d12} < P_{e12},\\
\frac{1}{P_{e12}}=\frac{1}{P_{d12}} & P_{e12} < P_{e2}, \{\theta_1 >1, \theta_2>1\},\\
&P_{d12} = P_{e12}.
\end{cases}
\end{equation}

\textit{Remark 3 (Optimization of the Violation Probability of AoI and AoA):}
It can be observed that the gradients of (\ref{viol_aoi}) and (\ref{viol_aoa_infi}) behave in the same way as the gradients of (\ref{AoI_mean}) and (\ref{AF_b_inf}) do.\footnote{We refrain from bringing the calculations and relations due to their size and repetitiveness in the logical process.} Hence, by similar arguments as in sections \ref{Age of Information_opt} and \ref{Optimization of the Age of Actuation}, the optimal violation probability of AoI and AoA are (\ref{Opt_points_AoI}) and (\ref{Opt_point_Ave_AoA}), respectively. Also, the optimal values can be calculated by inserting them into (\ref{viol_aoi}) and (\ref{viol_aoa}), respectively.

\subsection{Optimization of the Average AoI Constrained to the Average AoA} \label{Optimization of the Average AoI Constrained to the Average AoA}
We analyze the minimization of the average AoI constrained to the average AoA less than $\Delta$. The answer occurs in the energy-limited area. Thus, we solve $\bar{A}_2=\Delta$ for $q_1$ (or $q_2$), insert it into the average AoI, and then minimize it. The feasible critical point is
\begin{align*} 
\lambda_1=&\frac{P_{d1} P_{e2} (-P_{e12} + P_{e2}) \Delta}{P_{d1} (P_{e12} - P_{e2})^2 \Delta}\\
&-\frac{ \sqrt{P_{d1} (P_{d1} - P_{d12}) (P_{e12} - P_{e2})^2 P_{e2} \Delta}}{P_{d1} (P_{e12} - P_{e2})^2 \Delta} \numberthis \label{q1starofAACAA}
\end{align*}
and
\begin{equation}
\lambda_2=\frac{P_{d1} (-P_{e12} + P_{e2})}{\sqrt{P_{d1} (P_{d1} - P_{d12}) (P_{e12} -  P_{e2})^2 P_{e2} \Delta}}. 
\end{equation}
When $P_{e12} > P_{e2}$, or $P_{e12} < P_{e2}$ and (\ref{q1starofAACAA}) is above $1$, we have $q_1^*=1$, and thus $[q_1^*,q_2^*]=[1,\frac{1}{P_{e12} \Delta}]=[\phi_1,\phi_2]$, and its optimum value would be $\bar{I}_{\phi}=\frac{P_{e12} \Delta}{P_{d12} + P_{d1} ( P_{e12} \Delta -1)}$. When $P_{e12} < P_{e2}$ and (\ref{q1starofAACAA}) is below $1$, the optimum point is $[q_1^*,q_2^*]=[\lambda_1,\lambda_2]$, with the optimum value $\bar{I}_{\lambda}=\bar{I}(\lambda_1,\lambda_2)$. It can be summarized as
\begin{equation} \label{Solu_AoIconsAoA}
[q_1^*,q_2^*,\bar{I}^*]=
\begin{cases}
[\phi_1,\phi_2,\bar{I}_{\phi}] & P_{e12} > P_{e2}, \\
[\phi_1,\phi_2,\bar{I}_{\phi}] & P_{e12} < P_{e2}, q_1^* \geq 1,\\
[\lambda_1,\lambda_2,\bar{I}_{\lambda}] & P_{e12} < P_{e2}, q_1^* < 1.
\end{cases}
\end{equation}

\section{Probability of Missing the Actuation (PoMA)}
In this section, we characterize the probability of missing the actuation (PoMA) to capture when an action fails to occur for which a data packet has been generated. This can be associated with the potential cost due to that miss, thus the \textit{Cost of Missing Actuation} and \textit{Probability of Missing Actuation} can be used interchangeably here. Furthermore, we consider the optimization of PoMA given an average AoI constraint.

\subsection{The Analysis of PoMA}
Missing an actuation refers to the situation in which an action is issued from the transmitter (in the form of a data packet), and the expected actuation, based on the data, is missed. This missing can occur in two scenarios. Either the receiver fails to receive the data packet, or the data packet is received, but the system has no energy packet to support the actuation. Thus, at each time slot, the probability of missing an actuation is given by
\begin{equation} \label{Average_CoMA}
\bar{C}=q_1 q_2 \bar{P}_{d12}+q_1 \bar{q}_2 \bar{P}_{d1}+ 
q_1 q_2 P_{d12} \bar{P}_{e12} \pi_0 +q_1 \bar{q}_2 P_{d1} \pi_0.
\end{equation}
The first two terms are related to the first scenario, and the second two are related to the second scenario. 
From (\ref{empty_p_inf}) and (\ref{Average_CoMA}), we see that the probability of missing an actuation for the infinite-sized battery is
\begin{equation} \label{Average_CoMA_infinite}
\bar{C}=
\begin{cases}
\bar{C}_1=q_1 q_2 \bar{P}_{d12}+q_1 \bar{q}_2 \bar{P}_{d1} & \frac{P_{\bar{{\cal{D}}},{\cal{E}}}}{P_{{\cal{D}},\bar{{\cal{E}}}}} \geq 1,\\
\bar{C}_2=q_1 q_2 \bar{P}_{d12}+q_1 \bar{q}_2 \bar{P}_{d1}\\
+\left(1- \frac{P_{\bar{{\cal{D}}},{\cal{E}}}}{P_{{\cal{D}},\bar{{\cal{E}}}}}\right) \left(
q_1 q_2 P_{d12} \bar{P}_{e12}  +q_1 \bar{q}_2 P_{d1}\right) & \frac{P_{\bar{{\cal{D}}},{\cal{E}}}}{P_{{\cal{D}},\bar{{\cal{E}}}}} < 1.
\end{cases}
\end{equation}
The probability of missing an actuation for finite-sized batteries is
\begin{align*} \label{Average_CoMA_finite}
\bar{C}=&q_1 q_2 \bar{P}_{d12}+q_1 \bar{q}_2 \bar{P}_{d1}+ \left(\frac{1-\frac{P_{\bar{{\cal{D}}},{\cal{E}}}}{P_{{\cal{D}},\bar{{\cal{E}}}}}}{1-\left(\frac{P_{\bar{{\cal{D}}},{\cal{E}}}}{P_{{\cal{D}},\bar{{\cal{E}}}}}\right)^{m+1}}\right)\\
&\times \left( 
q_1 q_2 P_{d12} \bar{P}_{e12} +q_1 \bar{q}_2 P_{d1} \right) .
\end{align*}

\textit{Remark 4 (Energy Packet Drop Rate):} In the finite-sized battery, when the battery is full, and there is a successful transmission of an energy packet and an unsuccessful transmission of a data packet, the energy packet is dropped. Because the energy packet can be neither consumed nor stored, hence, its probability of occurrence at each time slot is $\bar{D}=\pi_{m}^{f} P_{\bar{{\cal{D}}},{\cal{E}}}$.

\subsection{Optimization of the PoMA Constrained to Average AoI} \label{Optimization of the Cost of Missing Actuation}
We will consider the optimization problem for the infinite-sized battery scenario. Similar to the average AoA, the function of (\ref{Average_CoMA_infinite}) has two pieces. The gradient vector of the first piece is
\begin{equation}
\nabla\bar{C}_1=[q_{2}\bar{P}_{d12} + \bar{q}_2 \bar{P}_{d1} , q_1\bar{P}_{d12} - q_1 \bar{P}_{d1}].
\end{equation}
Both elements are always positive. Its critical point, $[0,\frac{-1}{P_{d1}-P_{d12}}]$, is not in the valid domain, i.e., $0 \leq q_1, q_2 \leq 1$. For the second piece, the gradient of the objective function is
\begin{equation}
\nabla\bar{C}_2=[1+ q_{2}(P_{e2} - P_{e12}), - (1-q_1) P_{e12} - q_1 P_{e12}].
\end{equation}
The first element is always positive and the second element is always negative. The critical point is $[\frac{1}{P_{e12}-P_{e2}},\frac{P_{e2}}{P_{e2}-P_{e12}}]$, which cannot be in the square domain. Hence, there is no critical point in the domain. The intersection of these two surfaces, as we know, is (\ref{Equ_Border}). Opposite to optimizing the average AoA, here, the intersection of this line with the square domain at $q_1=q_2=0$ minimizes the probability of missing the actuation to zero, which is the minimum point. However, we want the average AoI to be lower than a threshold, i.e., $\bar{I}=\frac{1}{P_{\cal{D}}} \leq \Delta$. This requires non-zero $q_1$.

We investigate the optimum points on each area of $\bar{C}_1$ and $\bar{C}_2$. Both of these areas are decreasing on $q_1$. Hence, the minimum point occurs on the equality of the constraint, $\frac{1}{P_{\cal{D}}} = \Delta$. We investigate the minimum point of the line $\frac{1}{P_{\cal{D}}} = \Delta$ on each of the areas of $\bar{C}_1$ and $\bar{C}_2$.

For $\bar{C}_1$, we have $\bar{C}_1=q_1 - \Delta^{-1}$, for which there is no critical value of $q_1$ and to have lower value, $q_1$ should be as less as possible. That would be bounded by the intersection of the two areas. Hence, points of this type always occur at this intersection. Therefore, the intersection of the line $\frac{P_{\bar{{\cal{D}}},{\cal{E}}}}{P_{{\cal{D}},\bar{{\cal{E}}}}} = 1$ with the line $\bar{I}=\frac{1}{\Delta}$ should be investigated. Solving these two equations we obtain
\begin{equation} \label{Opt_Sulo_MA}
[q_1^*,q_2^*]=\Bigg[\frac{S + R - P_{d1} P_{e2} \Delta }
 {2 P_{d1} (P_{e12} - P_{e2}) \Delta},\frac{S - R + P_{d1} P_{e2} \Delta }
 {2 P_{e2} (P_{d1} - P_{d12}) \Delta}\Bigg] 
\end{equation}
in which $S=P_{d1} - P_{d12} + P_{e12} - P_{e2}$ and $R=\sqrt{ ( P_{d1} P_{e2} \Delta -S)^2 -4 P_{d1} P_{e2} (P_{e2}-P_{e12}) \Delta }$. Substituting (\ref{Opt_Sulo_MA}) in (\ref{Average_CoMA_infinite}), we have 
\begin{equation} \label{Opt_Sulo_MA_val}
\bar{C}^{*}=\frac{ ( S - R + P_{e2} \Delta (2-P_{d1})) (S + R -  P_{d1}  P_{e2} \Delta ) }{4 P_{d1}P_{e2} \Delta^2 ( P_{e12} - P_{e2}) }.
\end{equation}
Now, we investigate the minimum for when it occurs on $\bar{C}_2$. The minimum is obtained by minimizing the line of the equality constraint $\frac{1}{P_{\cal{D}}} = \Delta$ on $\bar{C}_2$.
We have
\begin{align*}
\bar{C}_2=&\frac{P_{e2} ( q_{1}-1) ( P_{d1} q_{1} \Delta -1)}{(P_{d1} - P_{d12}) q_{1} \Delta}\\
&+\frac{  q_{1} (P_{e12} + (P_{d1} - P_{d12}) q_{1} \Delta - P_{d1} P_{e12} q_{1} \Delta)}{(P_{d1} - P_{d12}) q_{1} \Delta}. \numberthis 
\end{align*}
Minimizing it yields
\begin{equation} \label{q1forC2}
q_1^*=\sqrt{\frac{P_{e2}}{(-P_{d12} + P_{d1} (1 - P_{e12} + P_{e2})) \Delta}}
\end{equation}
and
\begin{align*} 
q_2^*&=\frac{P_{d1}^2 P_{e2} \Delta - P_{d1} P_{d12} P_{e2} \Delta  }{(P_{d1} - P_{d12})^2 P_{e2} \Delta}\\
&-\frac{  \sqrt{(P_{d1} - P_{d12})^2 P_{e2} (  P_{d1} (1 - P_{e12} + P_{e2})-P_{d12}) \Delta}}{(P_{d1} - P_{d12})^2 P_{e2} \Delta}. \numberthis \label{q2forC2}
\end{align*}

The minimum value can be obtained by inserting (\ref{q1forC2}) and (\ref{q2forC2}) into $\bar{C}_2$.\footnote{The expression is omitted since it is elaborated.}

Different values of $\Delta$ separate the two different sets of results. Let $\hat{\Delta}$ be the value above which the minimums occur on the intersection and under which the minimums occur on the area of $\bar{C}_2$. Then $\hat{\Delta}$ can be obtained by inserting (\ref{q1forC2}) and (\ref{q2forC2}) into (\ref{Equ_Border}) and solve for $\Delta$.

\section{Numerical Results} \label{Numerical and Simulation Results}
This section provides numerical evaluations of the presented analytical results.
We consider two different setups.
For the first setup, the parameters are as follows: $P_{tx,1}=10~\text{mW}$, $P_{tx,2}=1~\text{W}$, $P_N=-50~ \text{dBm}$, $\gamma_d=\gamma_e=-10~ \text{dB}$, $d_1=1~\text{m}$ and $d_2=2~\text{m}$, $\alpha_1=\alpha_2=4$, $\upsilon_1=\upsilon_2=1$, and $\rho=0.99$. We alter only the value for $d_2=1.5~\text{m}$ for the second setup. The success probabilities are presented in Table \ref{Table_Setups}.
\begin{table}[b]
\begin{center}
\caption{Success probabilities of data and energy transmission.}
\label{Table_Setups}
\scalebox{1}{
\begin{tabular}{c|c|c|c|c|}
  \cline{2-5}
&   $P_{d1}$ & $P_{d12}$ & $P_{e2}$ & $P_{e12}$   \\  \cline{1-5}
\multicolumn{1}{ |c|  }{\multirow{1}{*}{Setup 1} } & 1 & 0.62 & 0.20 & 0.23  \\ \cline{1-5}
\multicolumn{1}{ |c|  }{\multirow{1}{*}{Setup 2} } & 1 & 0.34 & 0.60 & 0.63  \\ \cline{1-5}
\end{tabular}
}
\end{center}
\end{table}

In the first/second setup, receiving a data packet is more/less likely than receiving an energy packet when both transmitters are active. \textit{Note that $P_{e12} \geq P_{e2}$ is feasible, with large values of $\rho$. This is because, by power splitting, we can also utilize energy from the first transmitter.} In the figures, we show the optimal point by a data tip. Furthermore, the region of the limited energy regime is included in a black frame.

\subsection{Average AoI and AoA}
Figs. \ref{First_set_AoI} and \ref{Second_Set_AoI} show the results of the average AoI, against $q_1$ and $q_2$, for the first and the second setup, respectively. These results are independent of the battery. The minimum points confirm that to have the minimum average AoI, status updating and energizing should be carried out at the highest and least frequency, respectively. As we see, for the second setup, there is more sensitivity to $q_2$ compared to the first setup since, for the second setup, when there is interference, the success probability of the data is lower than the success probability of energy.

\ifdefined\TwoCol
\begin{figure}[h]
\centerline{\includegraphics[scale=0.64]{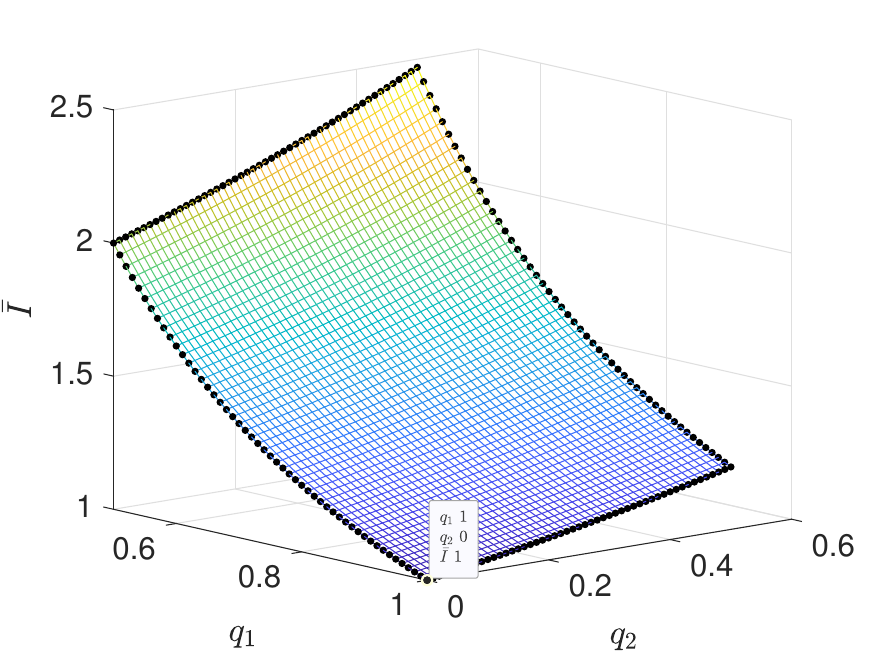}}
\caption{Average AoI for the first setup. The minimum $\bar{I}^*=1$ is achieved by $q^*_1=1$ and $q^*_2=0$.}
\label{First_set_AoI}
\end{figure}\
\begin{figure}[h]
\centerline{\includegraphics[scale=0.64]{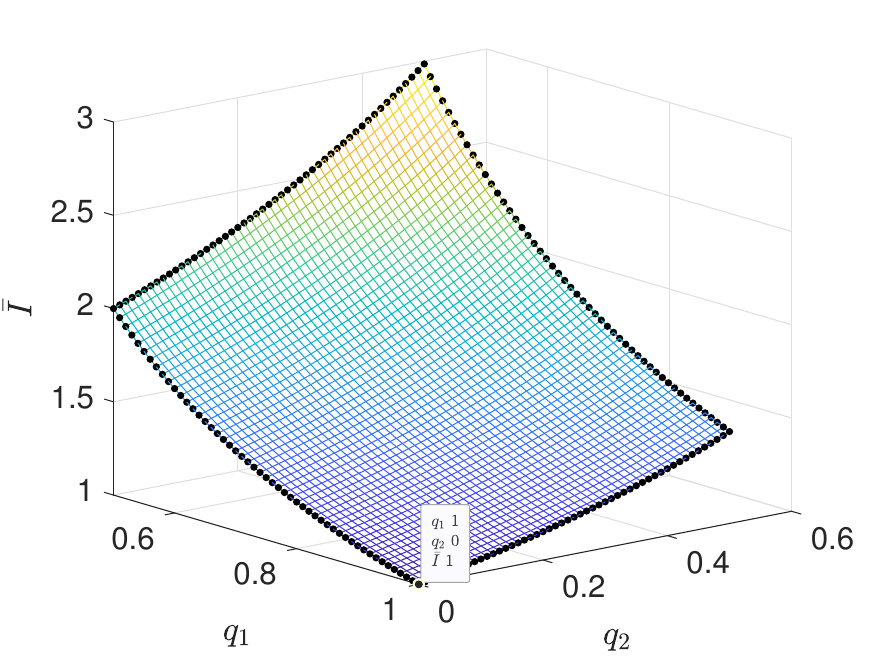}}
\caption{Average AoI for the second setup. The minimum $\bar{I}^*=1$ is achieved by $q^*_1=1$ and $q^*_2=0$.}
\label{Second_Set_AoI}
\end{figure}\
\fi
\ifdefined\OneCol
\begin{figure}[ht]
\centering
\begin{minipage}[b]{0.48\linewidth}
\centerline{\includegraphics[width=\textwidth]{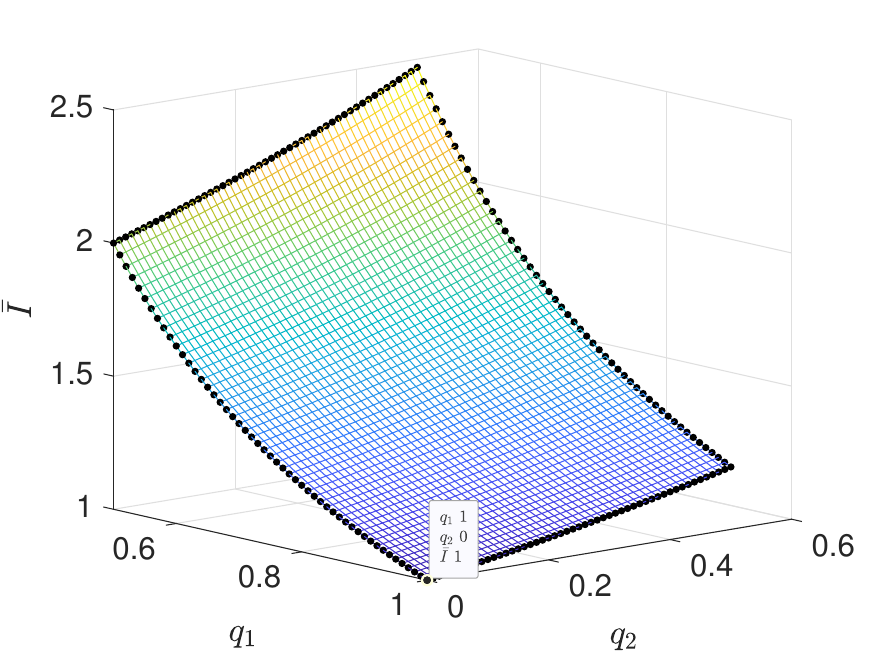}}
\caption{Average AoI for the first setup. The minimum $\bar{A}^*=1$ is achieved by $q^*_1=1$ and $q^*_2=0$.}
\label{First_Set_AoI}
\end{minipage}
\quad
\begin{minipage}[b]{0.48\linewidth}
\centerline{\includegraphics[width=\textwidth]{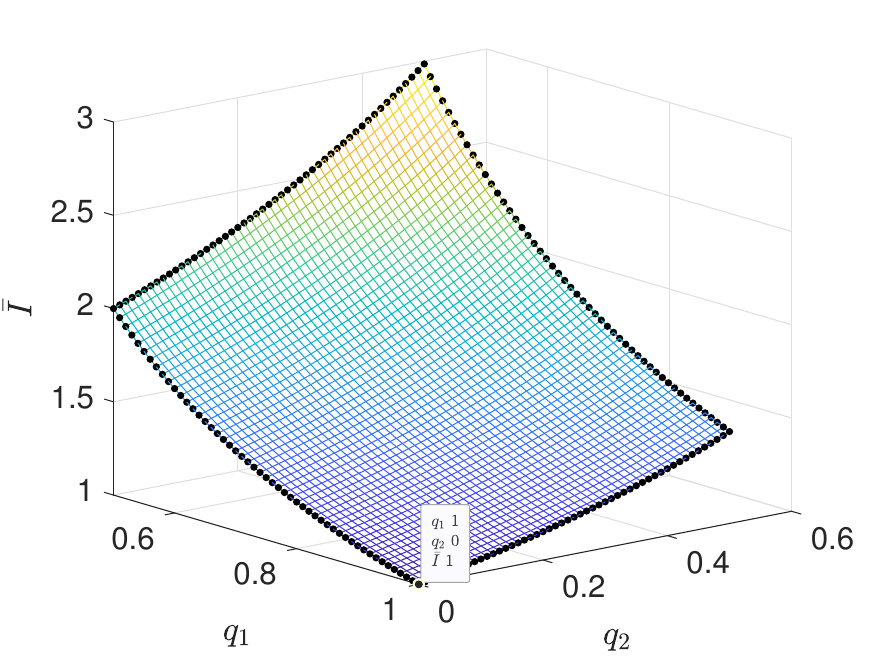}}
\caption{Average AoI for the second setup. The minimum $\bar{A}^*=1$ is achieved by $q^*_1=1$ and $q^*_2=0$.}
\label{Second_Set_AoI}
\end{minipage}
\end{figure}
\fi

Figs. \ref{First_set_AoA} and \ref{Second_Set_AoA} illustrate the average AoA for the case of an infinite-sized battery versus $q_1$ and $q_2$ for the first and the second setup, respectively. 
The minimum points and their values validate the analytical findings in (\ref{Opt_point_Ave_AoA}) and (\ref{Opt_value_Ave_AoA}). In Fig. \ref{First_set_AoA}, related to the first setup, we see that the minimum point is when $q^*_1=q^*_2=1$, which means it is the best to activate both transmitters.
\ifdefined\TwoCol
\begin{figure}[h]
\centerline{\includegraphics[scale=0.64]{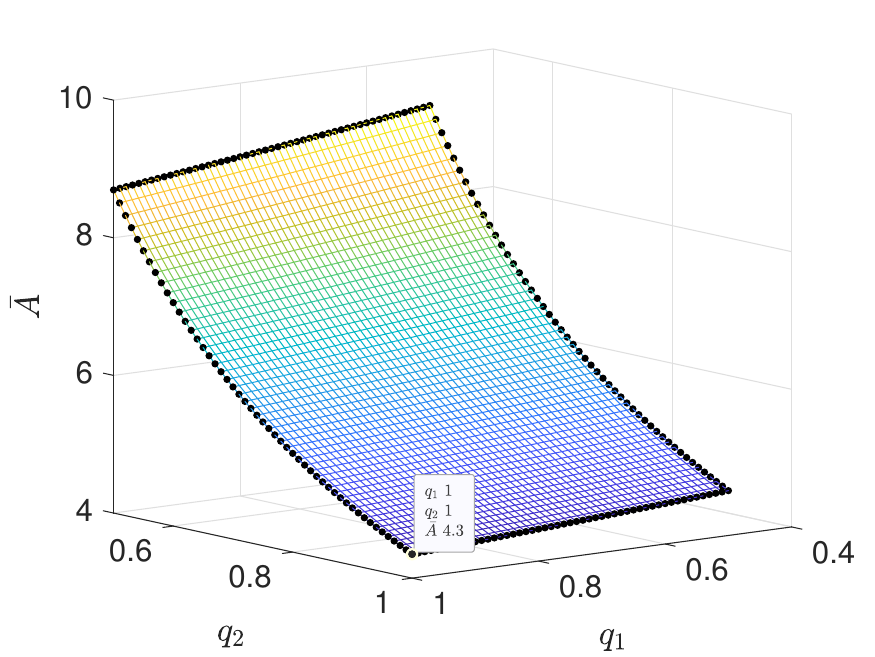}}
\caption{Average AoA for the infinite-sized battery for the first setup. The minimum $\bar{A}^{*}=4.3$ is achieved by $q^*_1=1$ and $q^*_2=1$.}
\label{First_set_AoA}
\end{figure}
\begin{figure}[h]
\centerline{\includegraphics[scale=0.64]{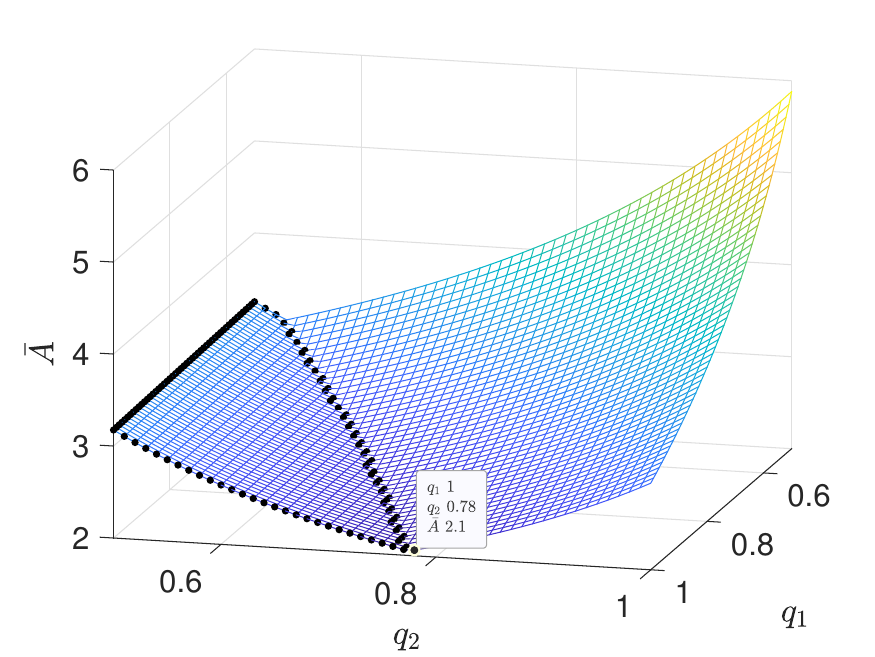}}
\caption{Average AoA for the infinite-sized battery for the second setup. The minimum $\bar{A}^{*}=2.1$ is achieved by $q^*_1=1$ and $q^*_2=0.78$.}
\label{Second_Set_AoA}
\end{figure}
\fi
\ifdefined\OneCol
\begin{figure}[ht]
\centering
\begin{minipage}[b]{0.48\linewidth}
\centerline{\includegraphics[scale=0.48]{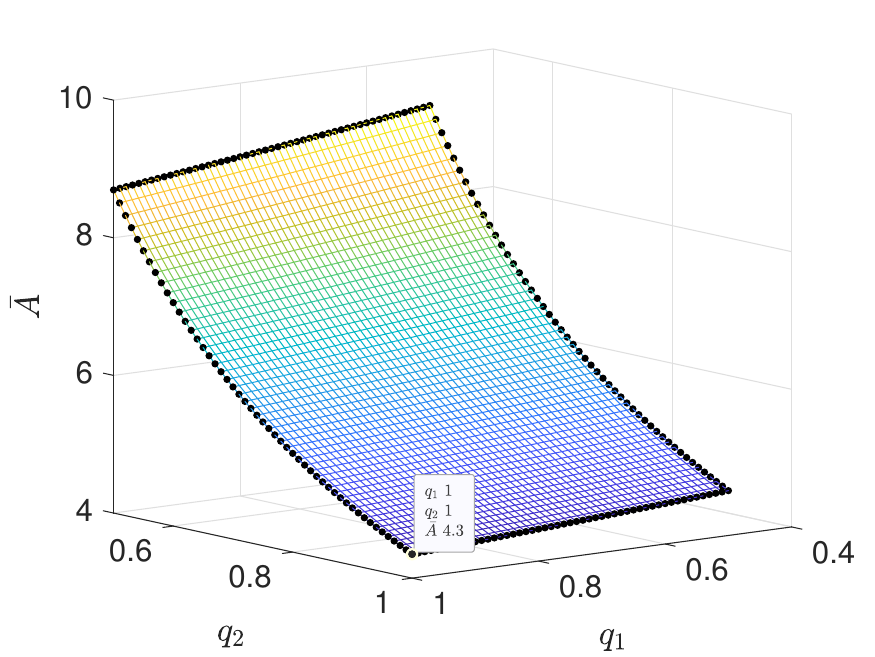}}
\caption{Average AoA for the infinite-sized battery for the first setup. The minimum $\bar{A}=4.3$ is achieved by $q^*_1=1$ and $q^*_2=1$.}
\label{First_set_AoA}
\end{minipage}
\quad
\begin{minipage}[b]{0.48\linewidth}
\centerline{\includegraphics[scale=0.48]{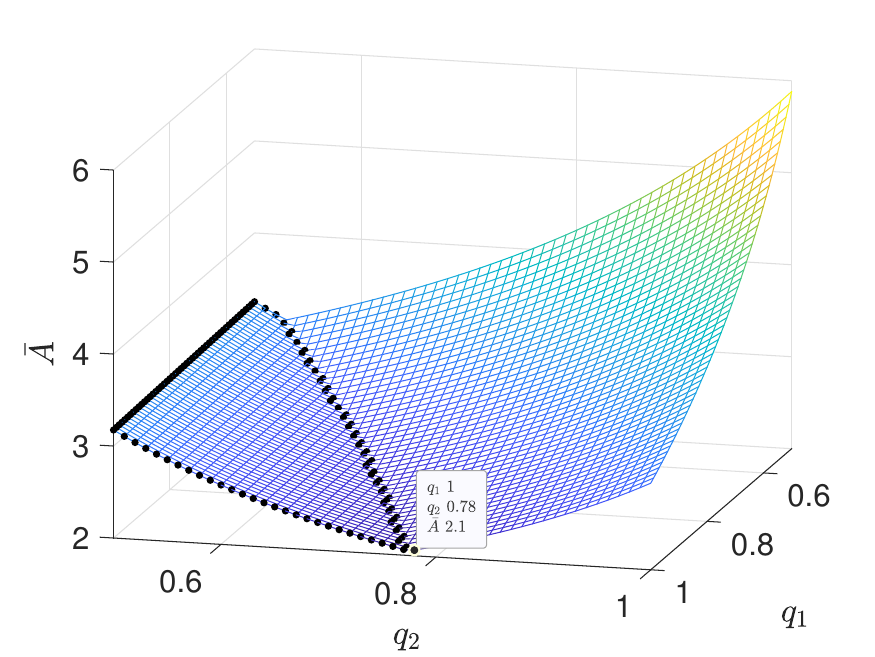}}
\caption{Average AoA for the infinite-sized battery for the second setup. The minimum $\bar{A}=2.1$ is achieved by $q^*_1=1$ and $q^*_2=0.78$.}
\label{Second_Set_AoA}
\end{minipage}
\end{figure}
\fi
In Fig. \ref{Second_Set_AoA}, related to the second setup in which interference can degrade the success probability for the data transmission significantly, we observe that $q^*_1=1$ and $q^*_2=0.78$. This is expected since we need to allow a period of silence for the transmitting power device not to degrade the transmission of the data node. Note that we cannot lower $q_2$ too much since we may receive data without sufficient energy to perform the actions.

In Figs. \ref{Finite1_First_set_AoA} and \ref{Finite1_Second_Set_AoA} we consider the case of having a battery of size $m=1$, for the first and the second setup, respectively. 
The minimum values for the finite-sized battery case are obtained by exhaustive search. Comparing Figs. \ref{Finite1_First_set_AoA} and \ref{First_set_AoA}, related to the first setup, there is no difference in the optimum point $[q^*_1, q^*_2]$. Therefore, \textit{from a design perspective, a small battery is sufficient for the optimal operation of that setup}. 

\ifdefined\TwoCol
\begin{figure}[h]
\centerline{\includegraphics[scale=0.64]{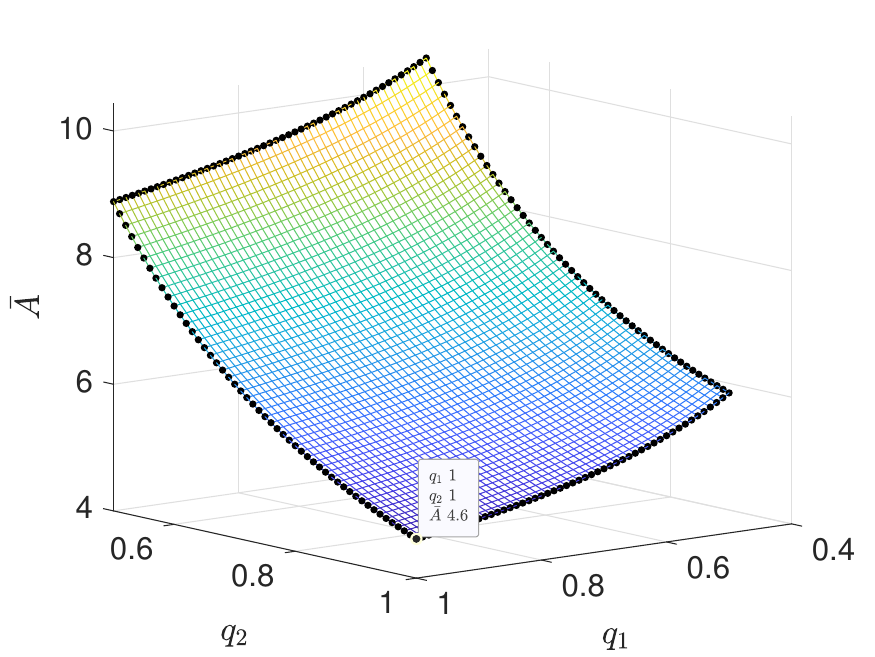}}
\caption{Average AoA for the finite-sized battery with the size of $m=1$, relate to the first setup. The minimum $\bar{A}^{*}=4.6$ is achieved by $q^*_1=q^*_2=1$.}
\label{Finite1_First_set_AoA}
\end{figure}
\begin{figure}[h]
\centerline{\includegraphics[scale=0.64]{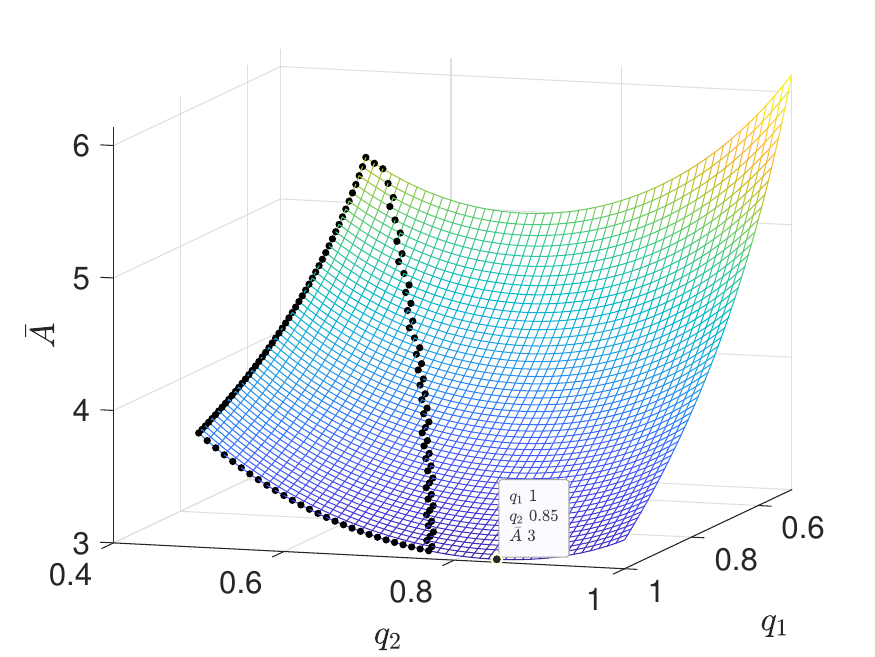}}
\caption{Average AoA for the finite-sized battery with the size of $m=1$, relate to the second setup. The minimum $\bar{A}^{*}=3$ is achieved by $q^*_1=1$ and $q^*_2=0.85$.}
\label{Finite1_Second_Set_AoA}
\end{figure}
\fi
\ifdefined\OneCol
\begin{figure}[ht]
\centering
\begin{minipage}[b]{0.48\linewidth}
\centerline{\includegraphics[scale=0.48]{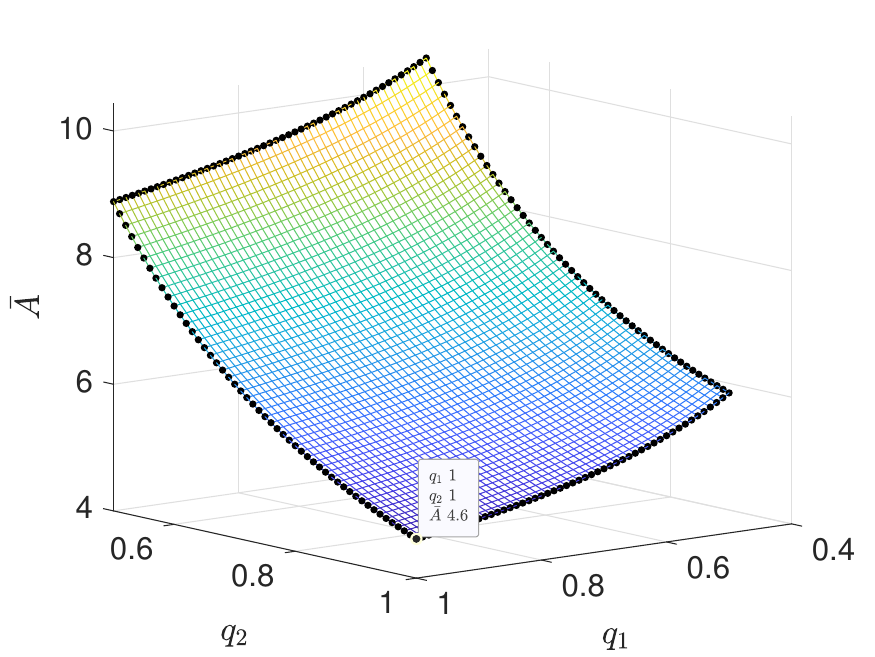}}
\caption{Average AoA for the finite-sized battery with the size of $m=1$, relate to the first setup. The minimum $\bar{A}=4.6$ is achieved by $q^*_1=q^*_2=1$.}
\label{Finite1_First_set_AoA}
\end{minipage}
\quad
\begin{minipage}[b]{0.48\linewidth}
\centerline{\includegraphics[scale=0.48]{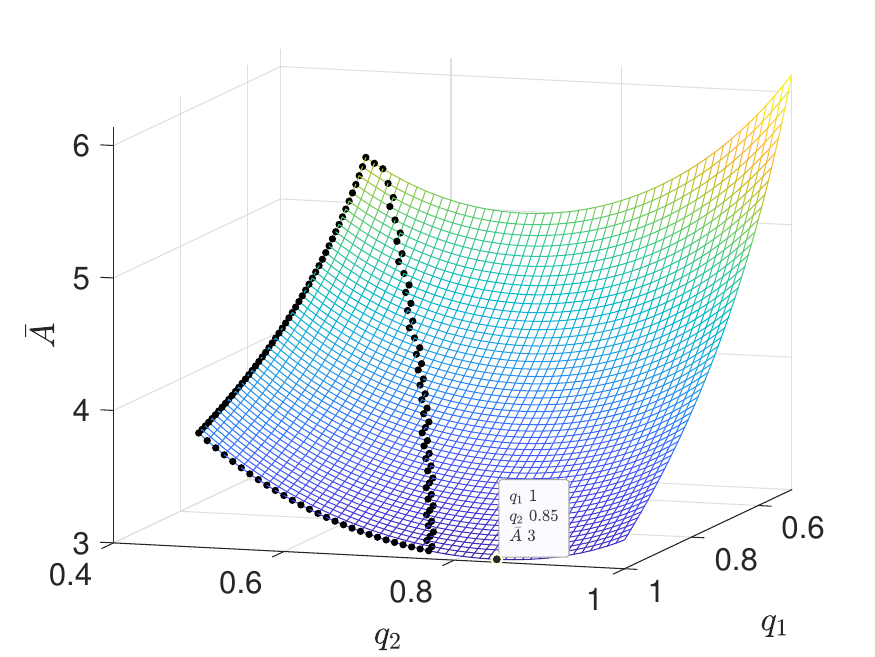}}
\caption{Average AoA for the finite-sized battery with the size of $m=1$, related to the second setup. The minimum $\bar{A}=3$ is achieved by $q^*_1=1$ and $q^*_2=0.85$.}
\label{Finite1_Second_Set_AoA}
\end{minipage}
\end{figure}
\fi

\ifdefined\TwoCol
\begin{figure}[h]
\centerline{\includegraphics[scale=0.64]{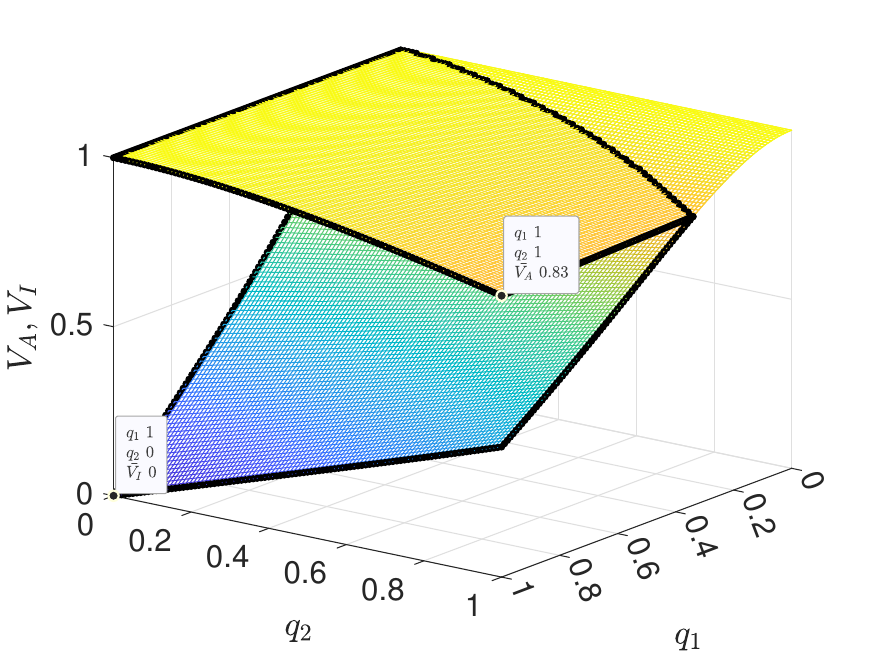}}
\caption{Violation probability of AoA and AoI for the infinite-sized battery for the first setup and $A_V=I_V=5$. The minimum $V_A^*=0.83$ is achieved by $q^*_1=1$ and $q^*_2=1$ for the AoA violation, and the minimum $V_I^*=0$ is achieved by $q^*_1=1$ and $q^*_2=0$ for the AoI violation.}
\label{First_Set_V}
\end{figure}
\begin{figure}[h]
\centerline{\includegraphics[scale=0.64]{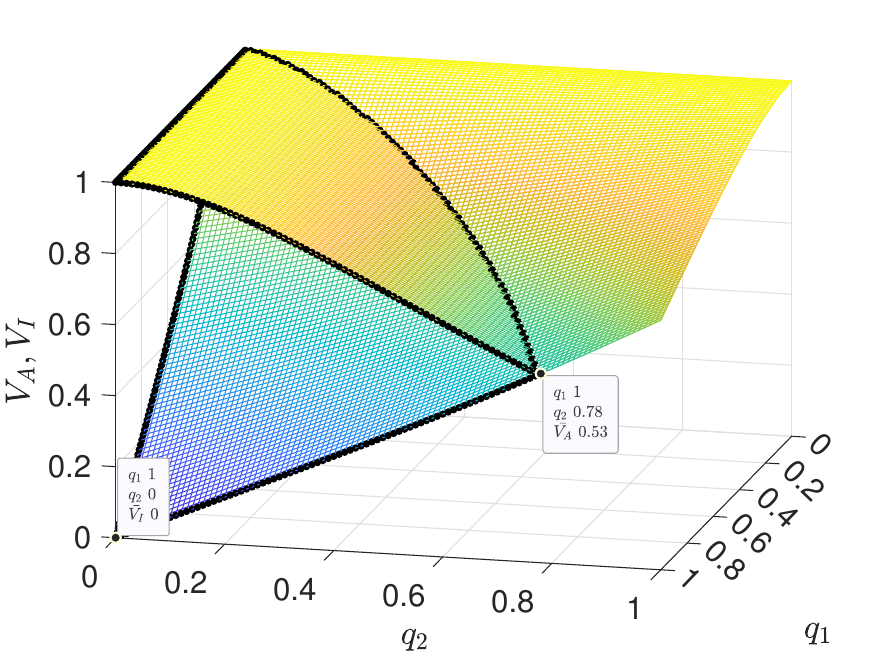}}
\caption{Violation probability of AoA and AoI for the infinite-sized battery for the second setup and $A_V=I_V=5$. The minimum $V_A^*=0.53$ is achieved by $q^*_1=1$ and $q^*_2=0.78$ for the AoA violation, and the minimum $V_I^*=0$ is achieved by $q^*_1=1$ and $q^*_2=0$ for the AoI violation.}
\label{Second_Set_V}
\end{figure}
\fi
\ifdefined\OneCol
\begin{figure}[ht]
\centering
\begin{minipage}[b]{0.48\linewidth}
\centerline{\includegraphics[scale=0.48]{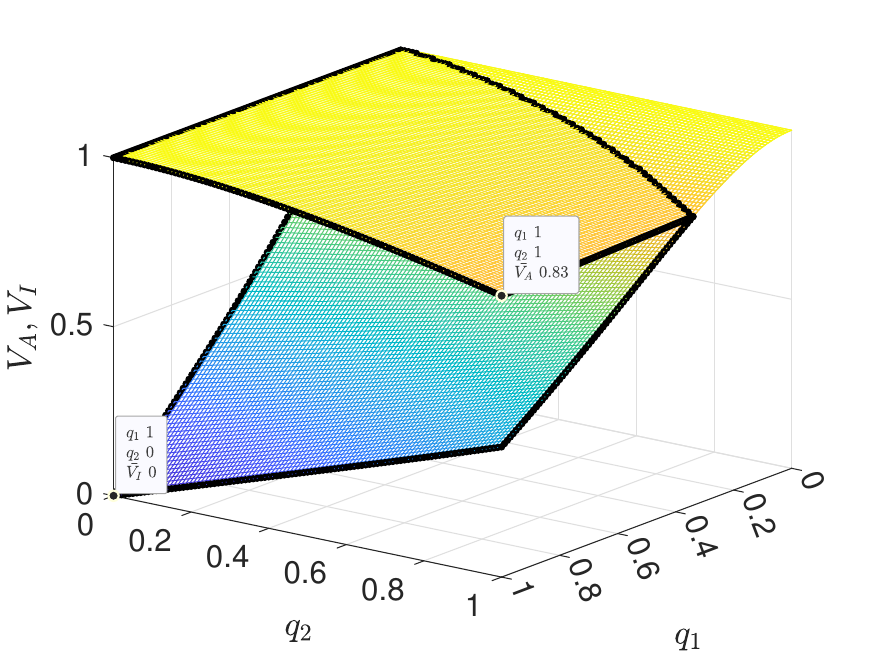}}
\caption{Violation probability of AoA and AoI for the infinite-sized battery for the first setup and $A_V=I_V=5$. The minimum $V_A^*=0.83$ is achieved by $q^*_1=1$ and $q^*_2=1$ for the AoA violation, and the minimum $V_I^*=0$ is achieved by $q^*_1=1$ and $q^*_2=0$ for the AoI violation.}
\label{First_Set_V}
\end{minipage}
\quad
\begin{minipage}[b]{0.48\linewidth}
\centerline{\includegraphics[scale=0.48]{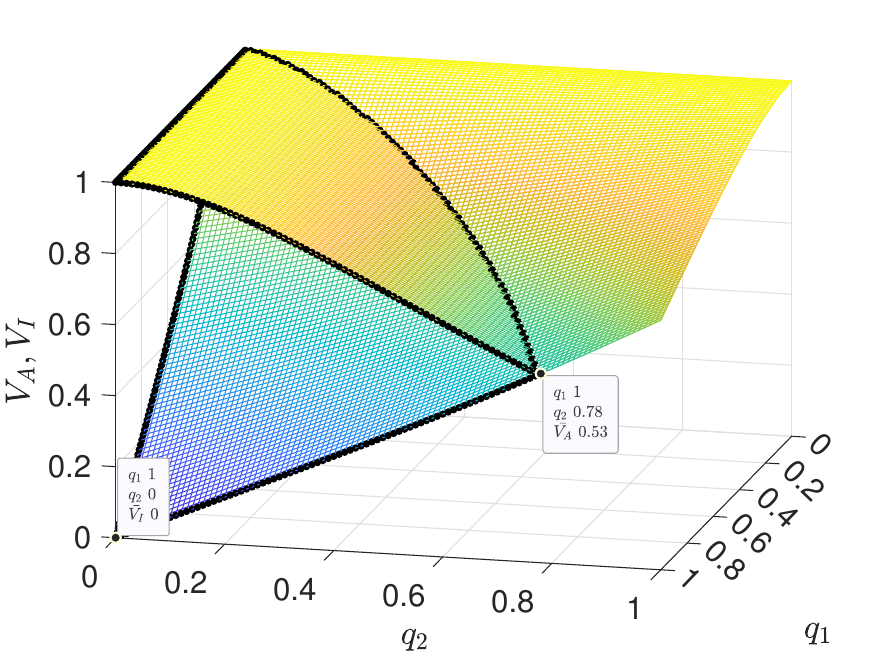}}
\caption{Violation probability of AoA and AoI for the infinite-sized battery for the second setup and $A_V=I_V=5$. The minimum $V_A^*=0.53$ is achieved by $q^*_1=1$ and $q^*_2=0.78$ for the AoA violation, and the minimum $V_I^*=0$ is achieved by $q^*_1=1$ and $q^*_2=0$ for the AoI violation.}
\label{Second_Set_V}
\end{minipage}
\end{figure}
\fi

\ifdefined\TwoCol
\begin{figure}[h]
\centerline{\includegraphics[scale=0.64]{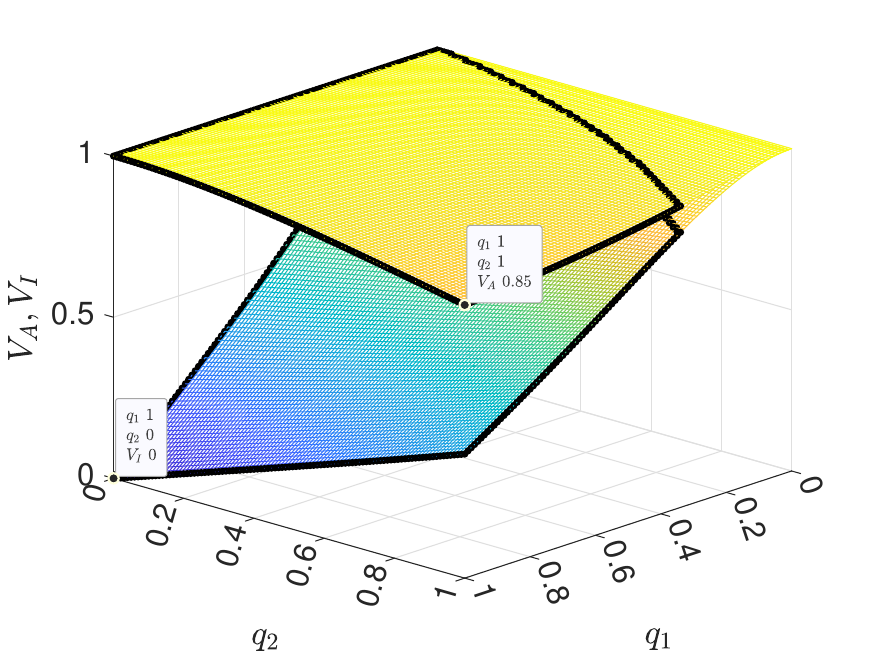}}
\caption{Violation probability of AoA and AoI for the finite-sized battery for the first setup and $A_V=I_V=5$. The minimum $V_A^*=0.85$ is achieved by $q^*_1=1$ and $q^*_2=1$ for the AoA violation, and the minimum $V_I^*=0$ is achieved by $q^*_1=1$ and $q^*_2=0$ for the AoI violation.}
\label{Finite1_First_Set_V}
\end{figure}
\begin{figure}[h]
\centerline{\includegraphics[scale=0.64]{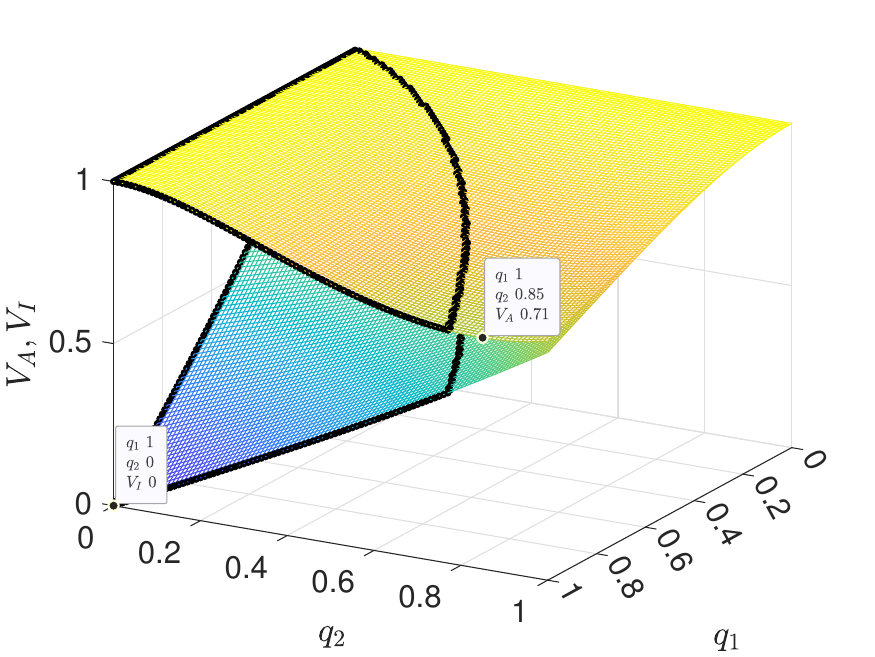}}
\caption{Violation probability of AoA and AoI for the finite-sized battery for the second setup and $A_V=I_V=5$. The minimum $V_A^*=0.71$ is achieved by $q^*_1=1$ and $q^*_2=0.85$ for the AoA violation, and the minimum $V_I^*=0$ is achieved by $q^*_1=1$ and $q^*_2=0$ for the AoI violation.}
\label{Finite1_Second_Set_V}
\end{figure}
\fi
\ifdefined\OneCol
\begin{figure}[ht]
\centering
\begin{minipage}[b]{0.48\linewidth}
\centerline{\includegraphics[scale=0.48]{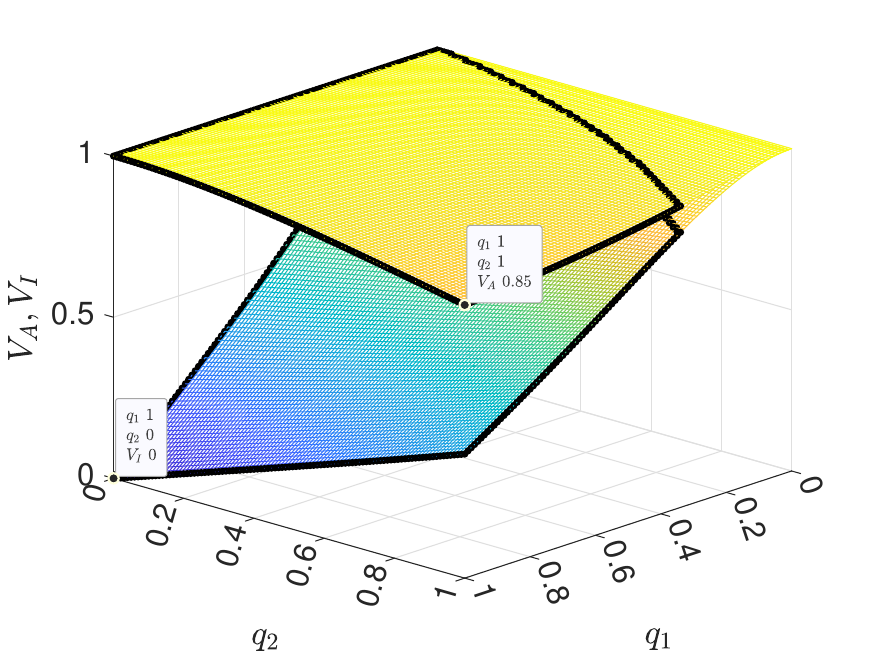}}
\caption{Violation probability of AoA and AoI for the finite-sized battery for the first setup and $A_V=I_V=5$. The minimum $V_A^*=0.85$ is achieved by $q^*_1=1$ and $q^*_2=1$ for the AoA violation, and the minimum $V_I^*=0$ is achieved by $q^*_1=1$ and $q^*_2=0$ for the AoI violation.}
\label{Finite1_First_Set_V}
\end{minipage}
\quad
\begin{minipage}[b]{0.48\linewidth}
\centerline{\includegraphics[scale=0.48]{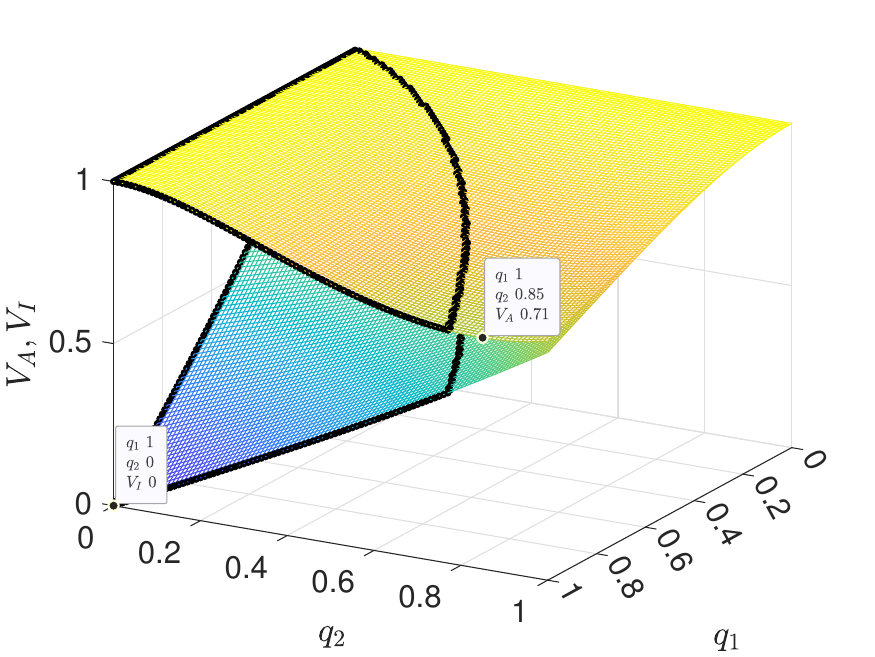}}
\caption{Violation probability of AoA and AoI for the finite-sized battery for the second setup and $A_V=I_V=5$. The minimum $V_A^*=0.71$ is achieved by $q^*_1=1$ and $q^*_2=0.85$ for the AoA violation and the minimum $V_I^*=0$ is achieved by $q^*_1=1$ and $q^*_2=0$ for the AoI violation.}
\label{Finite1_Second_Set_V}
\end{minipage}
\end{figure}
\fi

However, comparing Figs. \ref{Finite1_Second_Set_AoA} and \ref{Second_Set_AoA}, related to the second setup, we observe that as the battery size becomes smaller, the provision of energy is more critical. This is because the battery cannot store the energy packets due to the low capacity. Therefore, \textit{the battery may frequently be empty, and thus, many actuations have to occur by consuming energy packets that have just arrived}. This also has a significant impact on the minimum average AoA. 
As the battery size becomes larger, $q^*_2$ becomes smaller since the energy packets can be stored and utilized later. Asymptotically, the optimal point tends to the border of the energy-limited area.

\begin{table}
\begin{center}
\caption{Optimal points and values for different battery sizes.}
\label{Table_CoMA}
\begin{tabular}{cc|c|c|c|c|c|c|c|c|c|c|c|c|c|}
\cline{3-9}
& & \multicolumn{6}{ c| }{Finite Battery Size} & Inf. \\ \cline{2-9}
& \multicolumn{1}{ |c| }{$m$} & $1$  & $2$  & $3$ & $4$  & $5$  & $6$   & $\infty$ \\ \hline  \cline{1-9}
\multicolumn{1}{ |c|  }{Set. 1 } & $\bar{A}^*$ & $4.62$  & $4.40$    & $4.36$  & $4.35$  & $4.35$   & $4.35$ & $4.35$\\
 \hline 
 \cline{1-9}
\multicolumn{1}{ |c  }{\multirow{2}{*}{Set. 2} } &
\multicolumn{1}{ |c| }{$q_{2}^{*}$} & $0.85$  & $0.80$     & $0.79$  & $0.78$  & $0.78$  & $0.78$ & $0.78$ \\
\cline{2-9}
\multicolumn{1}{ |c  }{} &
\multicolumn{1}{ |c| }{$\bar{A}^*$} & $3.02$   & $2.62$ & $2.45$    & $2.36$  & $2.30$  & $2.26$ & $2.06$  \\ \cline{1-9}
\end{tabular}

\end{center}
\end{table}

As a comparison between the optimal points and values for infinite-sized and finite-sized batteries, table \ref{Table_CoMA} is presented. For the first setup, we always have $[q_{1}^{*},q_{2}^{*}]=[1,1]$, and for the second setup, we always have $q_{1}^{*}=1$.

\ifdefined\TwoCol
\begin{figure}[h]
\centerline{\includegraphics[scale=0.64]{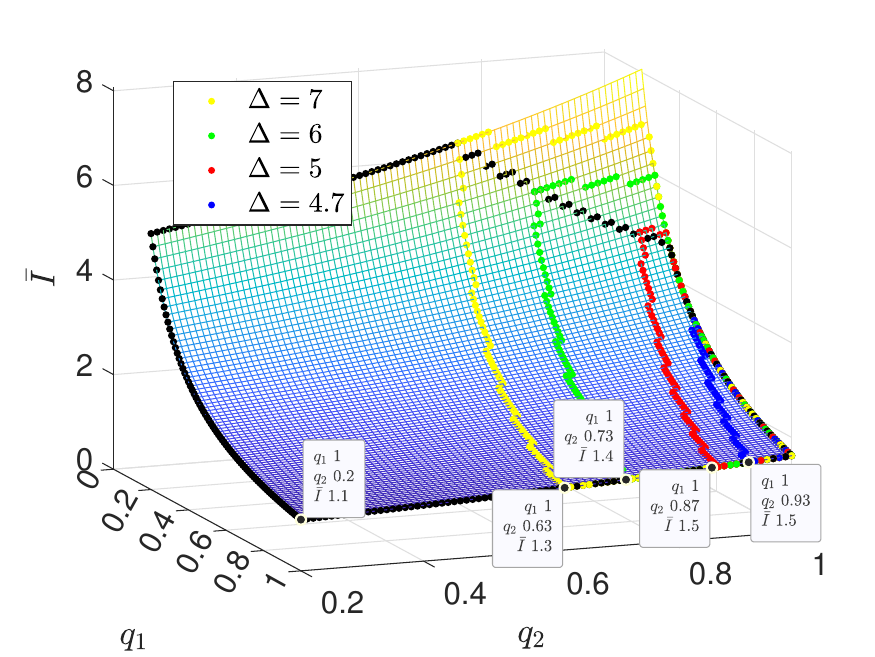}}
\caption{Average AoI constrained to the average AoA for the infinite-sized battery for the first setup. The minimums share the $q_1^*=1$. The minimum $\bar{I}^*=1.3$ is achieved by $q^*_1=1$ and $q^*_2=0.63$, $\bar{I}^*=1.4$ is achieved by $q^*_1=1$ and $q^*_2=0.73$, $\bar{I}^*=1.5$ is achieved by $q^*_1=1$ and $q^*_2=0.87$, and $\bar{I}^*=1.5$ is achieved by $q^*_1=1$ and $q^*_2=0.93$, for $\Delta=7$, $\Delta=6$, $\Delta=5$, and $\Delta=4.7$, respectively.}
\label{IcA_Inf_1}
\end{figure}
\begin{figure}[h]
\centerline{\includegraphics[scale=0.64]{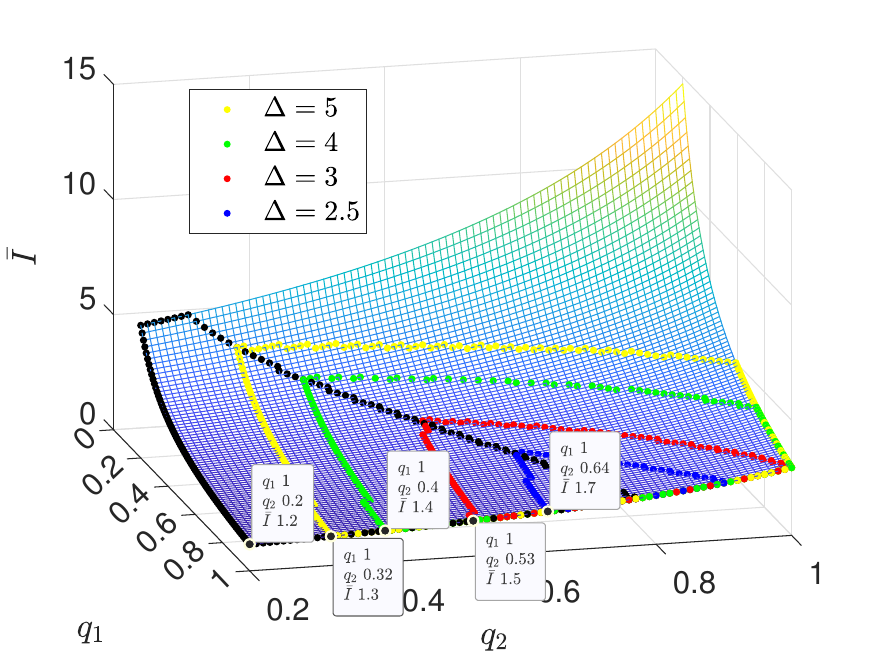}}
\caption{Average AoI constrained to the average AoA for the infinite-sized battery for the second setup. The minimums share the $q_1^*=1$. The minimum $\bar{I}^*=1.3$ is achieved by $q^*_1=1$ and $q^*_2=0.32$, $\bar{I}^*=1.4$ is achieved by $q^*_1=1$ and $q^*_2=0.4$, $\bar{I}^*=1.5$ is achieved by $q^*_1=1$ and $q^*_2=0.53$, and $\bar{I}^*=1.7$ is achieved by $q^*_1=1$ and $q^*_2=0.64$, for $\Delta=5$, $\Delta=4$, $\Delta=3$, and $\Delta=2.5$, respectively.}
\label{IcA_Inf_2}
\end{figure}
\fi
\ifdefined\OneCol
\begin{figure}[ht]
\centering
\begin{minipage}[b]{0.48\linewidth}
\centerline{\includegraphics[scale=0.48]{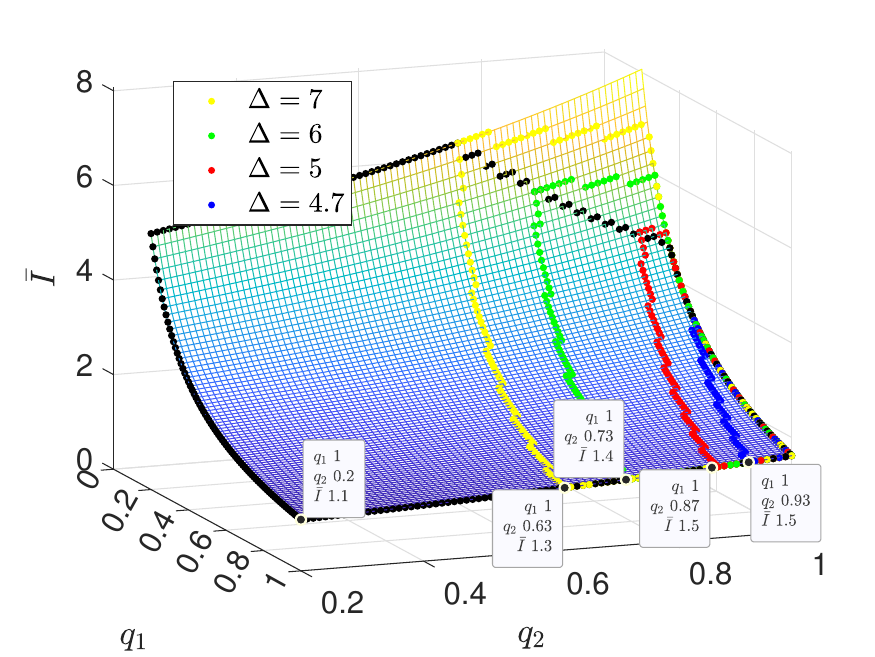}}
\caption{Average AoI constrained to the average AoA for the infinite-sized battery for the first setup. The minimums share the $q_1^*=1$. The minimums share the $q_1^*=1$. The minimum $\bar{I}^*=1.3$ is achieved by $q^*_1=1$ and $q^*_2=0.63$, $\bar{C}^*=1.4$ is achieved by $q^*_1=1$ and $q^*_2=0.73$, $\bar{C}^*=1.5$ is achieved by $q^*_1=1$ and $q^*_2=0.87$, and $\bar{C}^*=1.5$ is achieved by $q^*_1=1$ and $q^*_2=0.93$, for $\Delta=7$, $\Delta=6$, $\Delta=5$, and $\Delta=4.7$, respectively.}
\label{IcA_Inf_1}
\end{minipage}
\quad
\begin{minipage}[b]{0.48\linewidth}
\centerline{\includegraphics[scale=0.48]{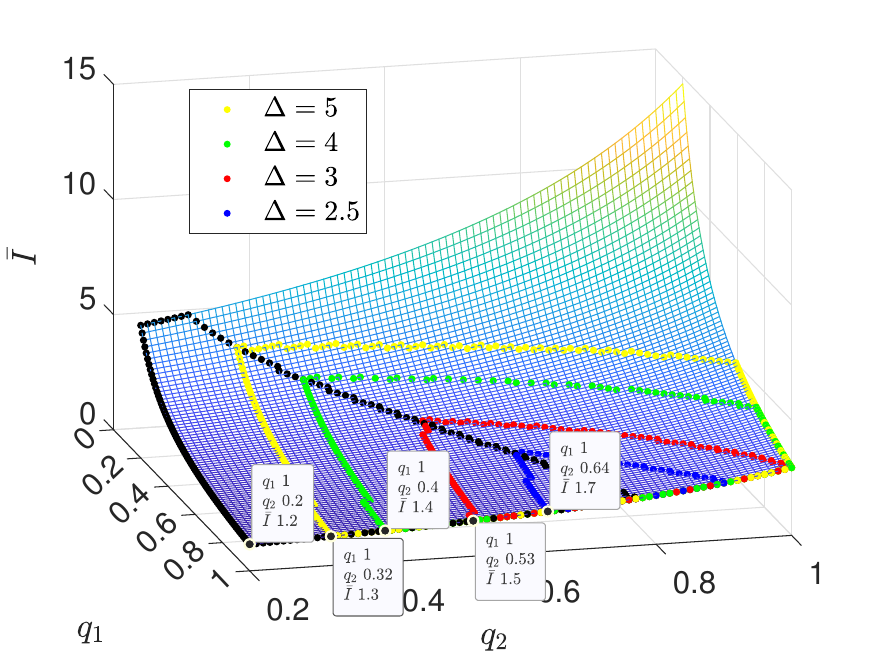}}
\caption{Average AoI constrained to the average AoA for the infinite-sized battery for the second setup. The minimums share the $q_1^*=1$. The minimum $\bar{I}^*=1.3$ is achieved by $q^*_1=1$ and $q^*_2=0.32$, $\bar{I}^*=1.4$ is achieved by $q^*_1=1$ and $q^*_2=0.4$, $\bar{I}^*=1.5$ is achieved by $q^*_1=1$ and $q^*_2=0.53$, and $\bar{I}^*=1.7$ is achieved by $q^*_1=1$ and $q^*_2=0.64$, for $\Delta=5$, $\Delta=4$, $\Delta=3$, and $\Delta=2.5$, respectively.}
\label{IcA_Inf_2}
\end{minipage}
\end{figure}
\fi
\ifdefined\TwoCol
\begin{figure}[h]
\centerline{\includegraphics[scale=0.64]{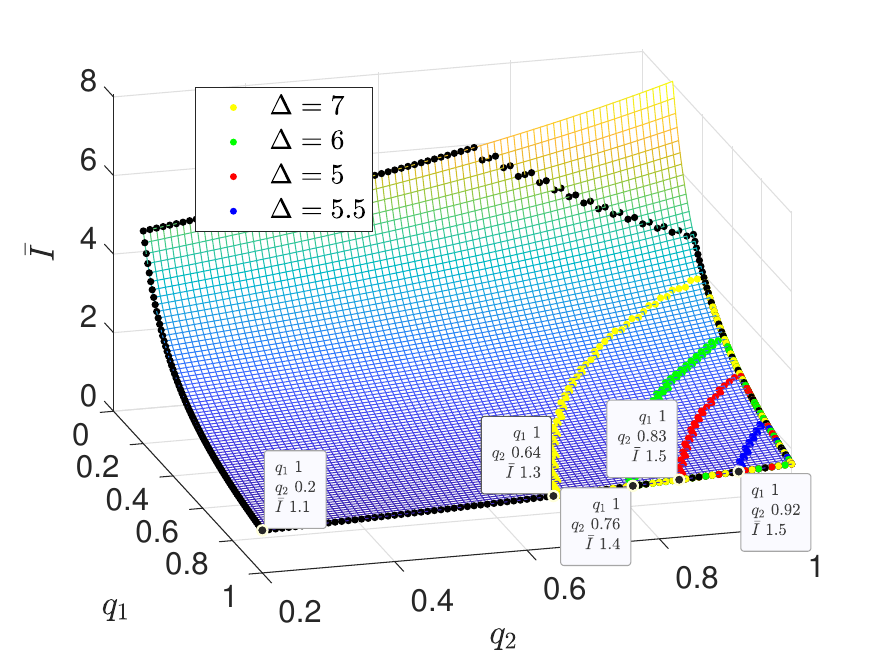}}
\caption{Average AoI constrained to the average AoA for the finite-sized battery, with size $m=1$, for the first setup. The minimums share the $q_1^*=1$. The minimum $\bar{I}^*=1.3$ is achieved by $q^*_1=1$ and $q^*_2=0.64$, $\bar{I}^*=1.4$ is achieved by $q^*_1=1$ and $q^*_2=0.76$, $\bar{I}^*=1.5$ is achieved by $q^*_1=1$ and $q^*_2=0.83$, and $\bar{I}^*=1.5$ is achieved by $q^*_1=1$ and $q^*_2=0.92$, for $\Delta=7$, $\Delta=6$, $\Delta=5$, and $\Delta=5.5$, respectively.}
\label{IcA_Fi_1}
\end{figure}
\begin{figure}[h]
\centerline{\includegraphics[scale=0.64]{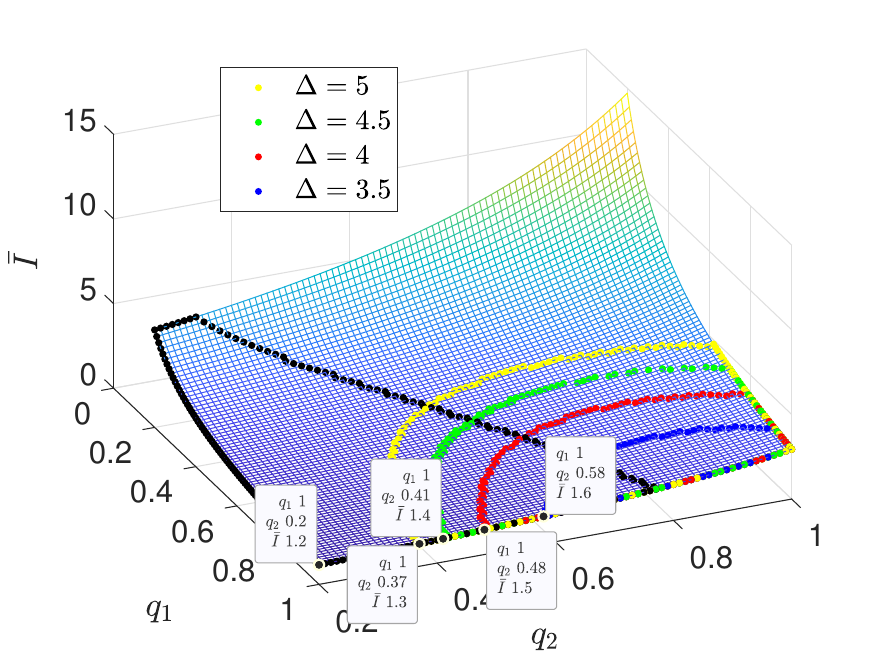}}
\caption{Average AoI constrained to the average AoA for the finite-sized battery, with size $m=1$, for the first setup. The minimums share the $q_1^*=1$. The minimum $\bar{I}^*=1.3$ is achieved by $q^*_1=1$ and $q^*_2=0.37$, $\bar{I}^*=1.4$ is achieved by $q^*_1=1$ and $q^*_2=0.41$, $\bar{I}^*=1.5$ is achieved by $q^*_1=1$ and $q^*_2=0.48$, and $\bar{I}^*=1.6$ is achieved by $q^*_1=1$ and $q^*_2=0.58$, for $\Delta=5$, $\Delta=4.5$, $\Delta=4$, and $\Delta=3.5$, respectively.}
\label{IcA_Fi_2}
\end{figure}
\fi
\ifdefined\OneCol
\begin{figure}[ht]
\centering
\begin{minipage}[b]{0.48\linewidth}
\centerline{\includegraphics[scale=0.48]{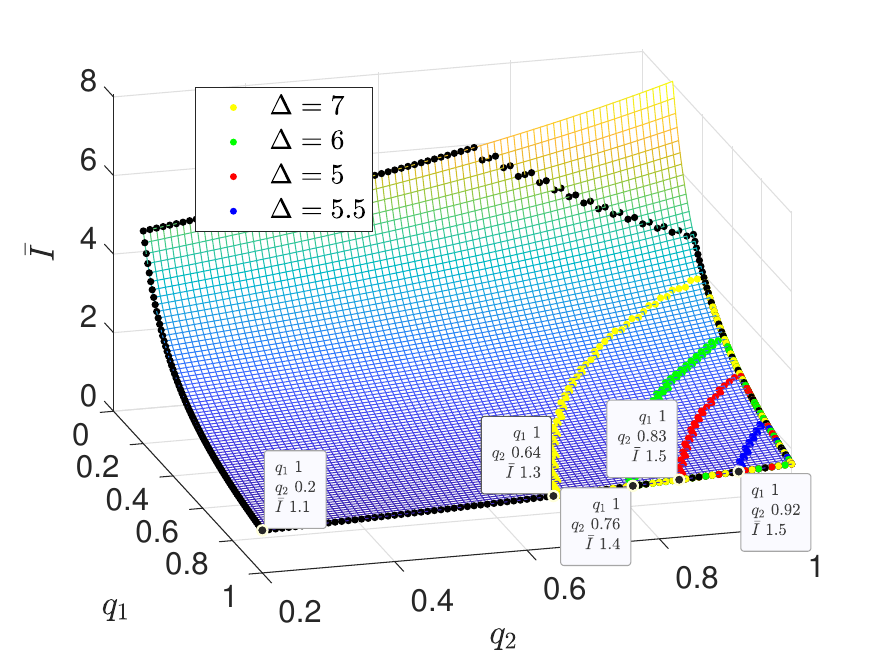}}
\caption{Average AoI constrained to the average AoA for the finite-sized battery, with size $m=1$, for the first setup. The minimums share the $q_1^*=1$. The minimum $\bar{I}^*=1.3$ is achieved by $q^*_1=1$ and $q^*_2=0.64$, $\bar{I}^*=1.4$ is achieved by $q^*_1=1$ and $q^*_2=0.76$, $\bar{I}^*=1.5$ is achieved by $q^*_1=1$ and $q^*_2=0.83$, and $\bar{I}^*=1.5$ is achieved by $q^*_1=1$ and $q^*_2=0.92$, for $\Delta=7$, $\Delta=6$, $\Delta=5$, and $\Delta=5.5$, respectively.}
\label{IcA_Fi_1}
\end{minipage}
\quad
\begin{minipage}[b]{0.48\linewidth}
\centerline{\includegraphics[scale=0.48]{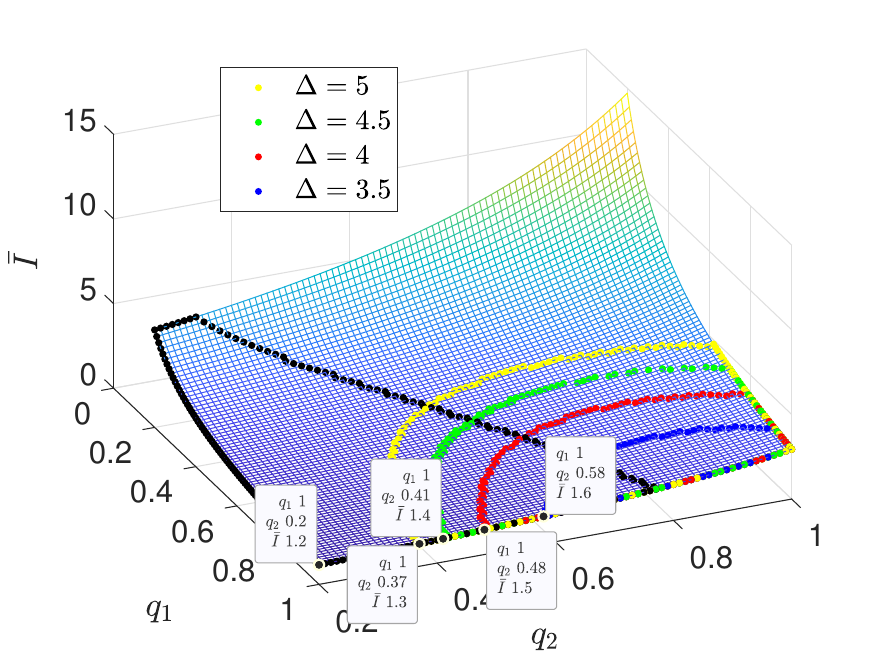}}
\caption{Average AoI constrained to the average AoA for the finite-sized battery, with size $m=1$, for the first setup. The minimums share the $q_1^*=1$. The minimum $\bar{I}^*=1.3$ is achieved by $q^*_1=1$ and $q^*_2=0.37$, $\bar{I}^*=1.4$ is achieved by $q^*_1=1$ and $q^*_2=0.41$, $\bar{I}^*=1.5$ is achieved by $q^*_1=1$ and $q^*_2=0.48$, and $\bar{I}^*=1.6$ is achieved by $q^*_1=1$ and $q^*_2=0.58$, for $\Delta=5$, $\Delta=4.5$, $\Delta=4$, and $\Delta=3.5$, respectively.}
\label{IcA_Fi_2}
\end{minipage}
\end{figure}
\fi

\subsection{Violation Probabilities of the Average AoI and the Average AoA}

To facilitate a comparison between the AoI and AoA, we simultaneously present the results of their violation probability when $V_A=V_I=5$. Figs. \ref{First_Set_V} and \ref{Second_Set_V} illustrate the corresponding outcomes for the first and second setups, assuming infinite-sized batteries. The areas of overlap and the areas above represent the violation probability of AoA, whereas the overlapped areas and areas beneath represent the violation probability of AoI. Figs. \ref{Finite1_First_Set_V} and \ref{Finite1_Second_Set_V} provide analogous results for finite-sized batteries, where the above and below areas are associated with the violation probabilities of AoA and AoI, respectively.

The minimum points for violation probability for every $V_I$ and $V_A$ are identical and coincide with the minimum points for the average AoI and AoA. This analytical alignment is elaborated upon in Remark 3. For further clarification, refer to Figs. \ref{First_set_AoI}, \ref{Second_Set_AoI}, \ref{First_set_AoA}, \ref{Second_Set_AoA}, \ref{Finite1_First_set_AoA}, and \ref{Finite1_Second_Set_AoA}.

\subsection{Average AoI Constrained to the Average AoA}
Figs. \ref{IcA_Inf_1} and \ref{IcA_Inf_2}, and \ref{IcA_Fi_1} and \ref{IcA_Fi_2} show the results of optimization of the average AoI constrained to the average AoA being less than $\Delta$, for the first and the second set of success probabilities for infinite-sized batteries, and the first and the second set of success probabilities for finite-sized batteries, respectively.

In this optimization problem, the feasible area shrinks as $\Delta$ decreases. Asymptotically, it will be a point, $\Delta_0$. For the threshold $\Delta_0$, the optimum point of this optimization is the same as the optimum point of the optimization of the average AoA. Also, $\Delta_0$ is the optimum value of the average AoA. Figs. \ref{First_set_AoA}, \ref{Second_Set_AoA}, \ref{Finite1_First_set_AoA}, and \ref{Finite1_Second_Set_AoA} can confirm the the explained result.

\subsection{PoMA Constrained to the Average AoI}

\ifdefined\TwoCol
\begin{figure}[h]
\centerline{\includegraphics[scale=0.64]{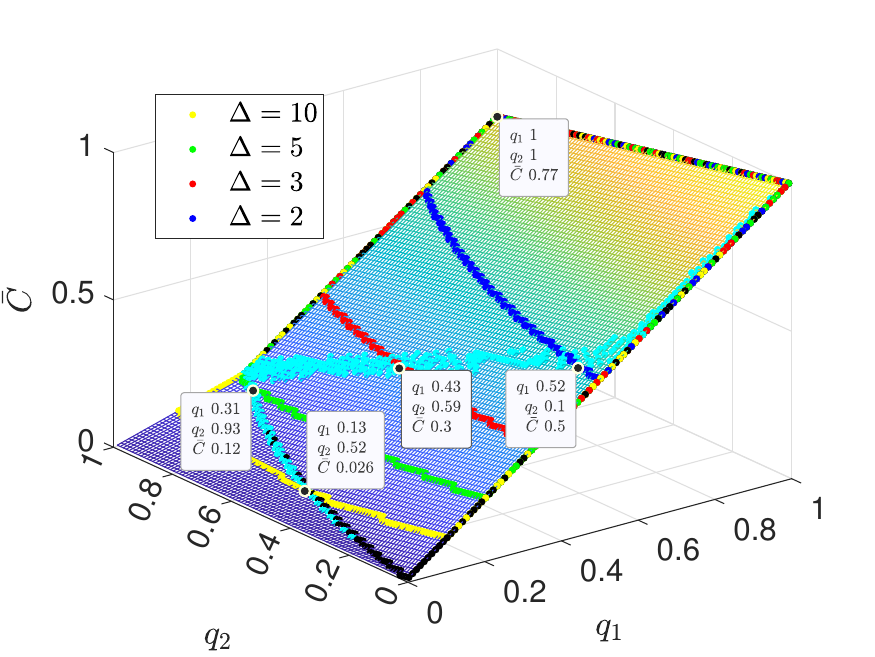}}
\caption{Average PoMA constrained to the average AoI for the infinite-sized battery for the first setup. The minimum $\bar{C}^*=0.5$ is achieved by $q^*_1=0.52$ and $q^*_2=0.1$, $\bar{C}^*=0.3$ is achieved by $q^*_1=0.43$ and $q^*_2=0.59$, $\bar{C}^*=0.12$ is achieved by $q^*_1=0.31$ and $q^*_2=0.93$, and $\bar{C}^*=0.026$ is achieved by $q^*_1=0.13$ and $q^*_2=0.52$, for $\Delta=2$, $\Delta=3$, $\Delta=5$, and $\Delta=10$, respectively.}
\label{First_set_CoA}
\end{figure}\
\begin{figure}[h]
\centerline{\includegraphics[scale=0.64]{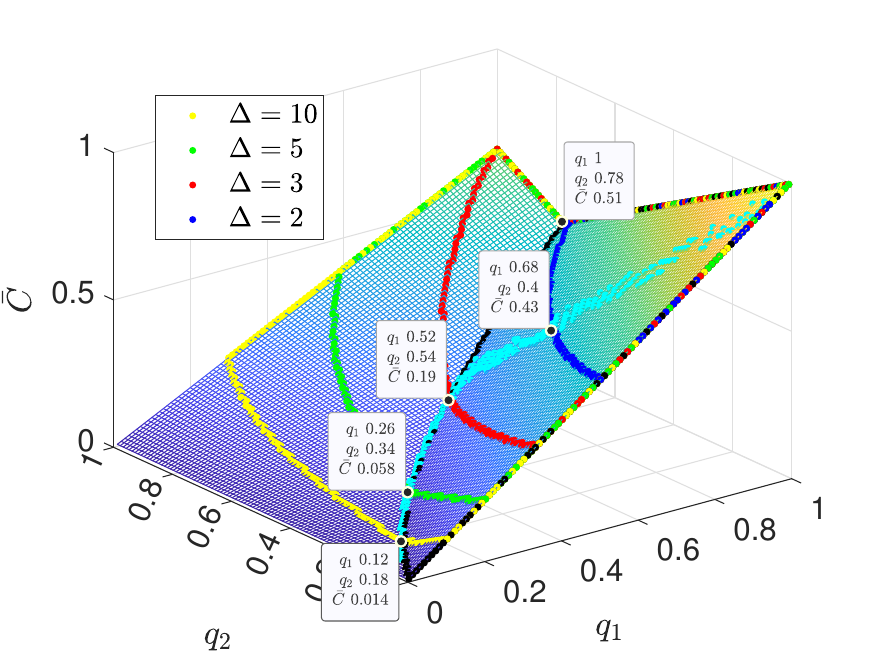}}
\caption{Average PoMA constrained to the average AoI for the infinite-sized battery for the first setup. The minimum $\bar{C}^*=0.43$ is achieved by $q^*_1=0.68$ and $q^*_2=0.4$, $\bar{C}^*=0.19$ is achieved by $q^*_1=0.52$ and $q^*_2=0.54$, $\bar{C}^*=0.058$ is achieved by $q^*_1=0.26$ and $q^*_2=0.34$, and $\bar{C}^*=0.014$ is achieved by $q^*_1=0.12$ and $q^*_2=0.18$, for $\Delta=2$, $\Delta=3$, $\Delta=5$, and $\Delta=10$, respectively.}
\label{Second_Set_CoA}
\end{figure}\
\fi
\ifdefined\OneCol
\begin{figure}[ht]
\centering
\begin{minipage}[b]{0.48\linewidth}
\centerline{\includegraphics[width=\textwidth]{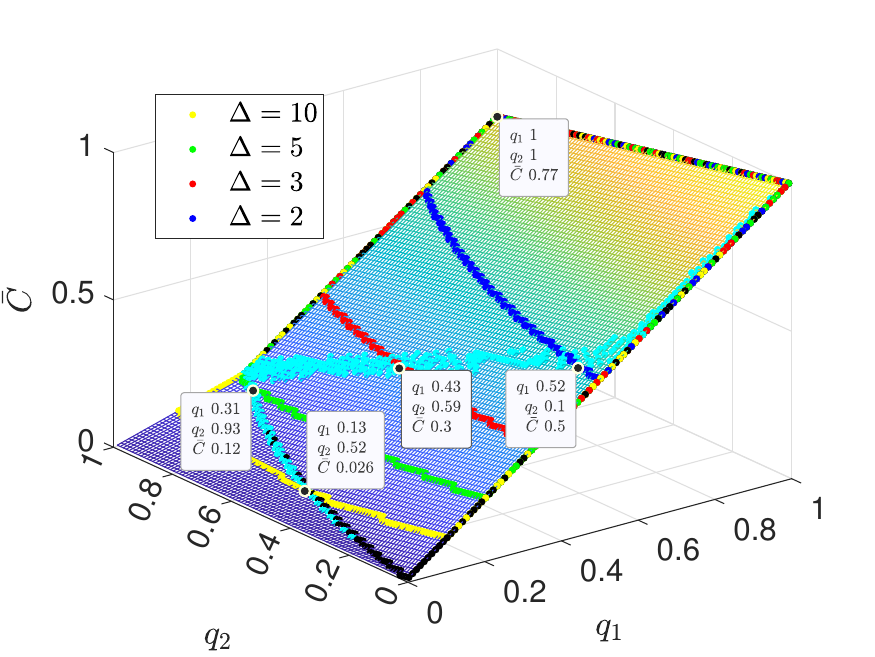}}
\caption{Average PoMA constrained to the average AoI for the infinite-sized battery for the first setup. The minimums $\bar{C}^*=0.5$ is achieved by $q^*_1=0.52$ and $q^*_2=0.1$, $\bar{C}^*=0.3$ is achieved by $q^*_1=0.43$ and $q^*_2=0.59$, $\bar{C}^*=0.12$ is achieved by $q^*_1=0.31$ and $q^*_2=0.93$, and $\bar{C}^*=0.026$ is achieved by $q^*_1=0.13$ and $q^*_2=0.52$, for $\Delta=2$, $\Delta=3$, $\Delta=5$, and $\Delta=10$, respectively.}
\label{First_Set_CoA}
\end{minipage}
\quad
\begin{minipage}[b]{0.48\linewidth}
\centerline{\includegraphics[width=\textwidth]{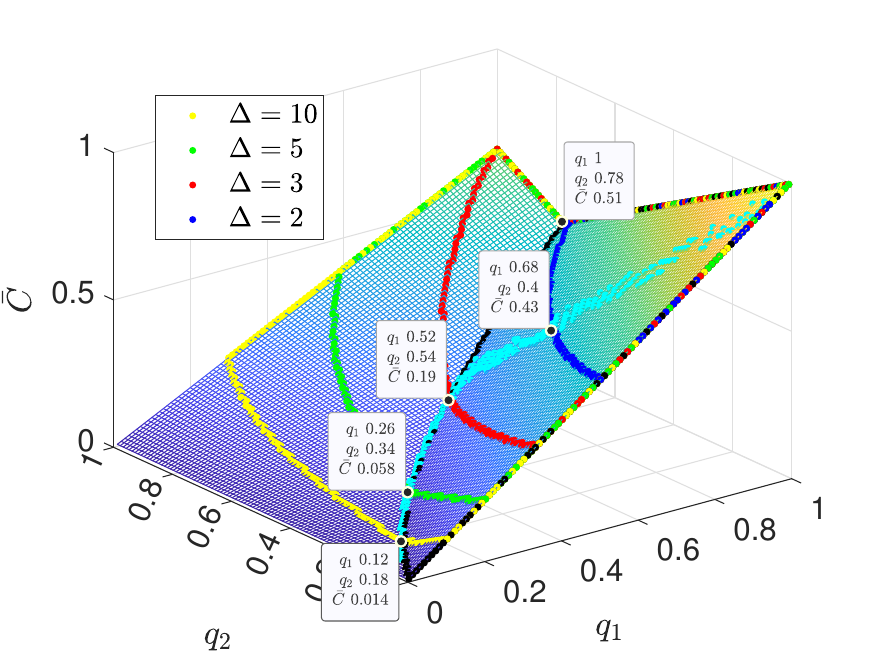}}
\caption{Average PoMA constrained to the average AoI for the infinite-sized battery for the first setup. The minimums $\bar{C}^*=0.43$ is achieved by $q^*_1=0.68$ and $q^*_2=0.4$, $\bar{C}^*=0.19$ is achieved by $q^*_1=0.52$ and $q^*_2=0.54$, $\bar{C}^*=0.058$ is achieved by $q^*_1=0.26$ and $q^*_2=0.34$, and $\bar{C}^*=0.014$ is achieved by $q^*_1=0.12$ and $q^*_2=0.18$, for $\Delta=2$, $\Delta=3$, $\Delta=5$, and $\Delta=10$, respectively.}
\label{Second_Set_CoA}
\end{minipage}
\end{figure}
\fi

We demonstrate the results of the average PoMA for the two different setups, presented in Table \ref{Table_Setups}. The areas where the condition of $\bar{I} \leq \Delta$ holds for $\Delta=10$, $\Delta=5$, $\Delta=3$, and $\Delta=2$ are confined within yellow, green, red, and blue frames, respectively. Also, the minimum for each of the four $\Delta$s is shown by a data tip. Besides, minimums related to values of $\Delta=\{1.01,1.02,\hdots,19.99,20.00\}$ are determined by cyan points. In addition to the minimum points of the average PoMA, the minimum of the average AoA is shown for the corresponding setup and the battery condition for each figure.

Figs. \ref{First_set_CoA} and \ref{Second_Set_CoA} show the average PoMA for the first and the second setup, respectively, for the infinite-sized battery. The results validate our findings in section \ref{Optimization of the Cost of Missing Actuation}. In fact, the minimums related to $\Delta=10$ and $\Delta=5$ in Fig. \ref{First_set_CoA}, and minimums related to $\Delta=10$, $\Delta=5$, and $\Delta=3$ in Fig. \ref{Second_Set_CoA} can be obtained from (\ref{Opt_Sulo_MA}) and (\ref{Opt_Sulo_MA_val}). Also the minimums related to the $\Delta=3$ and $\Delta=2$ in Fig. \ref{First_set_CoA}, and minimums related to $\Delta=2$ in Fig. \ref{Second_Set_CoA} can be obtained from (\ref{q1forC2}) and (\ref{q2forC2}) and its obtainable $\Bar{C}^*$. It is clear how they lay on the $\bar{C}_2$ area and the intersections of $\bar{C}_1$ and $\bar{C}_2$ areas. In Figs. \ref{First_set_CoA} and \ref{Second_Set_CoA}, when the average AoI constraint gets tighter the values of $q_1^*$ grow larger. Meanwhile, the values of $q_2^*$ are increased to the point that $\Delta=\Delta^*$, and then are decreased. This different behavior is more significant for the second setup, as shown in Figs. \ref{First_set_CoA} and \ref{Second_Set_CoA}. It is because when there is interference, in the first setup, data has a larger success probability of transmission than energy, and the optimization can tolerate larger $q_2$s.

\ifdefined\TwoCol
\begin{figure}[h]
\centerline{\includegraphics[scale=0.64]{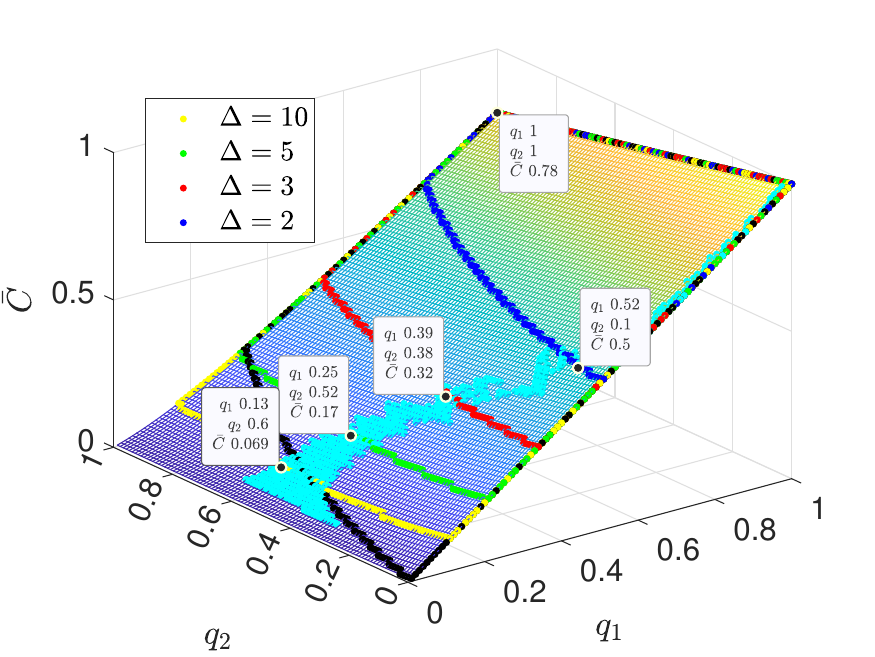}}
\caption{Average PoMA constrained to the average AoI for the finite-sized battery with the size of $m=1$, for the first setup. The minimum $\bar{C}^*=0.5$ is achieved by $q^*_1=0.52$ and $q^*_2=0.1$, $\bar{C}^*=0.32$ is achieved by $q^*_1=0.39$ and $q^*_2=0.38$, $\bar{C}^*=0.17$ is achieved by $q^*_1=0.25$ and $q^*_2=0.52$, and $\bar{C}^*=0.069$ is achieved by $q^*_1=0.13$ and $q^*_2=0.6$, for $\Delta=2$, $\Delta=3$, $\Delta=5$, and $\Delta=10$, respectively.}
\label{Finite1_First_set_CoA}
\end{figure}\
\begin{figure}[h]
\centerline{\includegraphics[scale=0.64]{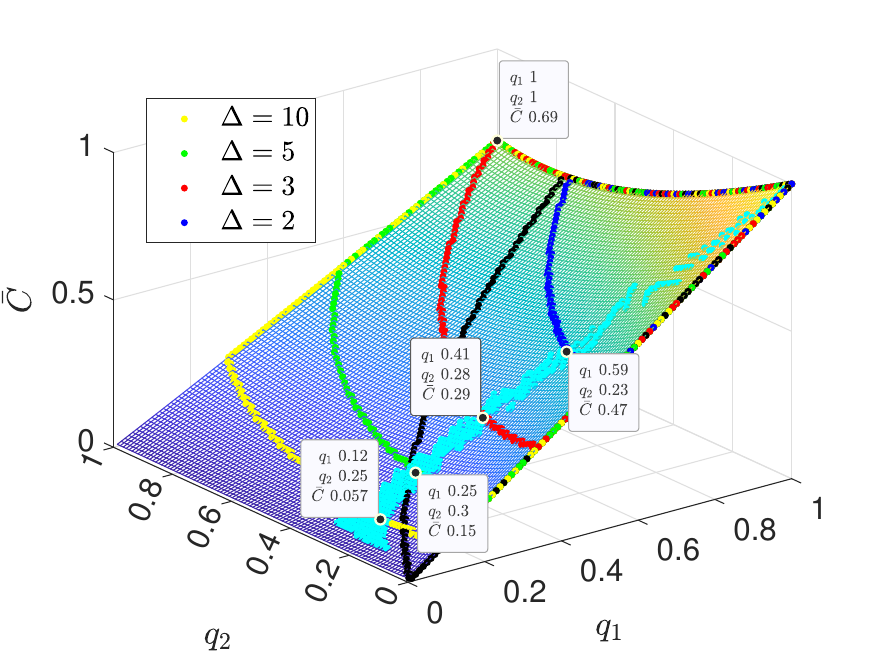}}
\caption{Average PoMA constrained to the average AoI for the finite-sized battery with the size of $m=1$, for the second setup. The minimum $\bar{C}^*=0.47$ is achieved by $q^*_1=0.59$ and $q^*_2=0.23$, $\bar{C}^*=0.29$ is achieved by $q^*_1=0.41$ and $q^*_2=0.28$, $\bar{C}^*=0.15$ is achieved by $q^*_1=0.25$ and $q^*_2=0.3$, and $\bar{C}^*=0.057$ is achieved by $q^*_1=0.12$ and $q^*_2=0.25$, for $\Delta=2$, $\Delta=3$, $\Delta=5$, and $\Delta=10$, respectively.}
\label{Finite1_Second_Set_CoA}
\end{figure}\
\fi
\ifdefined\OneCol
\begin{figure}[ht]
\centering
\begin{minipage}[b]{0.48\linewidth}
\centerline{\includegraphics[width=\textwidth]{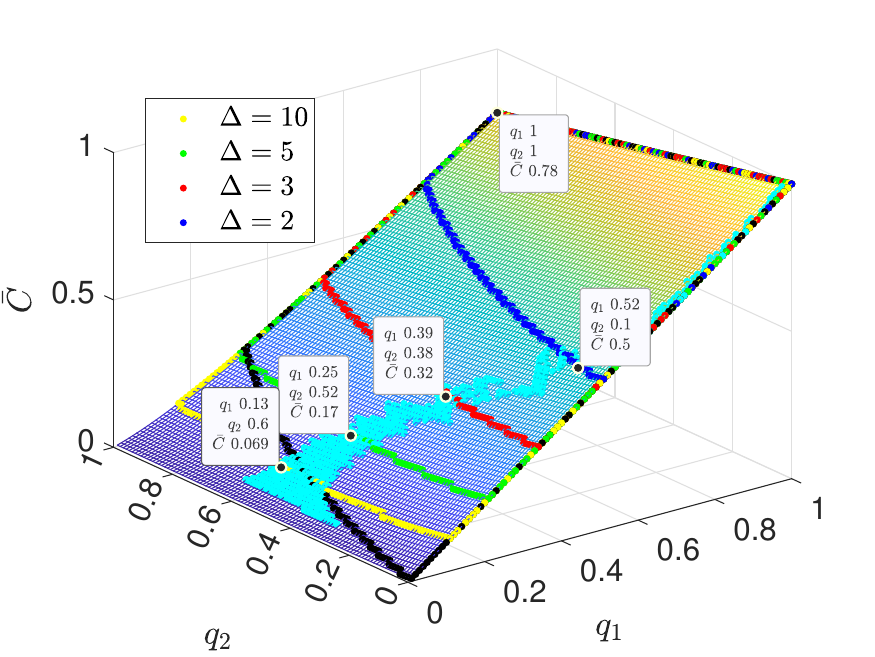}}
\caption{Average PoMA constrained to the average AoI for the finite-sized battery with the size of $m=1$, for the first setup. The minimum $\bar{C}^*=0.5$ is achieved by $q^*_1=0.52$ and $q^*_2=0.1$, $\bar{C}^*=0.32$ is achieved by $q^*_1=0.39$ and $q^*_2=0.38$, $\bar{C}^*=0.17$ is achieved by $q^*_1=0.25$ and $q^*_2=0.52$, and $\bar{C}^*=0.069$ is achieved by $q^*_1=0.13$ and $q^*_2=0.6$, for $\Delta=2$, $\Delta=3$, $\Delta=5$, and $\Delta=10$, respectively.}
\label{Finite1_First_Set_CoA}
\end{minipage}
\quad
\begin{minipage}[b]{0.48\linewidth}
\centerline{\includegraphics[width=\textwidth]{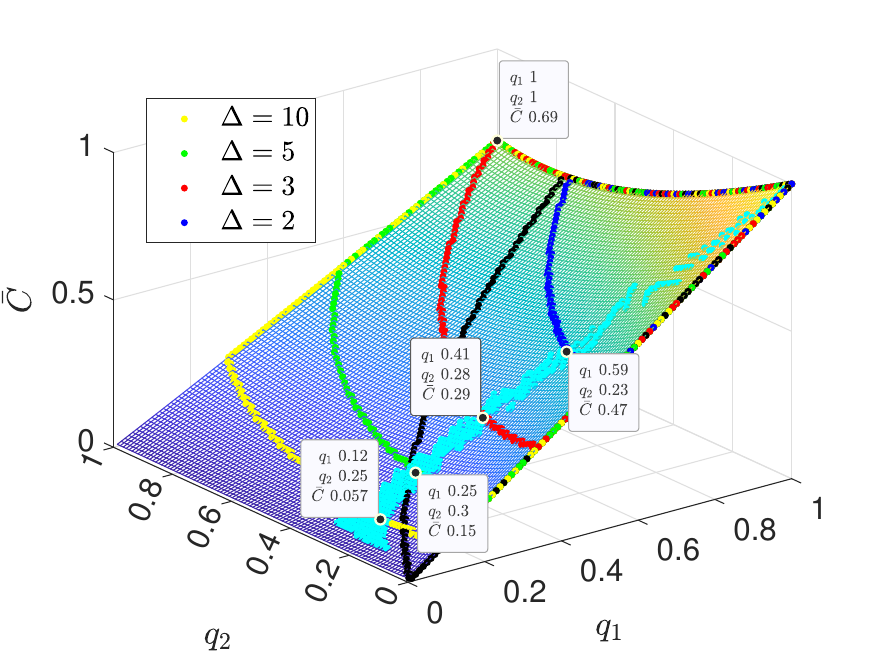}}
\caption{Average PoMA constrained to the average AoI for the finite-sized battery with the size of $m=1$, for the second setup. The minimum $\bar{C}^*=0.47$ is achieved by $q^*_1=0.59$ and $q^*_2=0.23$, $\bar{C}^*=0.29$ is achieved by $q^*_1=0.41$ and $q^*_2=0.28$, $\bar{C}^*=0.15$ is achieved by $q^*_1=0.25$ and $q^*_2=0.3$, and $\bar{C}^*=0.057$ is achieved by $q^*_1=0.12$ and $q^*_2=0.25$, for $\Delta=2$, $\Delta=3$, $\Delta=5$, and $\Delta=10$, respectively.}
\label{Finite1_Second_Set_CoA}
\end{minipage}
\end{figure}
\fi

We demonstrate the results of the average PoMA for the battery size of $m=1$ for the two setups in Figs. \ref{Finite1_First_set_CoA} and \ref{Finite1_Second_Set_CoA}. We can see that when the battery is highly limited in capacity, the $q_1^*$ increases with tightening the constraint of the average AoI, while the $q_2^*$ slowly decreases.

\section{Conclusion}
In this study, we investigated a system with a transmitter responsible for monitoring a process and transmitting status updates to a receiver. These updates inform the receiver about the process status and enable actuation. The receiver, powered by a battery, is charged using a secondary transmitter dedicated to wireless energy transmission, with occasional assistance from the primary transmitter. We studied the performance of such a system in terms of Age of Information (AoI). We defined a new metric that becomes relevant when actions based on status updates are performed: Age of Actuation (AoA), which offers a broader perspective than AoI. We optimized both the average AoI and the average AoA. Moreover, we optimized the average AoI constrained to the average AoA. Tightening the constraint, the asymptotic feasible point of this optimization is the optimum point of the average AoA, and the value of the optimum average AoA is the asymptotic threshold of this optimization. We characterized the violation probabilities of AoI and AoA and showed that their optimum points are the same as the averages of the two metrics. In addition, we defined another metric called the probability of missing the actuation (PoMA) that is relevant when command data is generated and sent from the transmitter and related actuation is performed. We solved its optimization, given that a minimum average AoI is guaranteed.

Our findings shed light on the system's performance characteristics and optimization opportunities. The results emphasize the importance of balancing the timeliness of status updates (AoI) and the actuation (AoA), as well as considering the probability of missing actuation (PoMA).

\section*{Appendix A}
We have $P_{e12}=\text{Pr}\left\{\rho^2\left(P_{rx,1} +  P_{rx,2}\right) \geq \gamma_e \right\}$. We calculate
\begin{equation} \label{1minusT_e12_App}
1-P_{e12}=\text{Pr}\left\{ \left(g_1 A_1 + g_2 A_2\right) \leq \frac{\gamma_e }{ \rho^2 } \right\} .
\end{equation}
The probability density function (PDF) of $A$ \cite{nguyen2008optimization} multiplied by a constant $g > 0$, i.e., $gA$, is 
\begin{equation}  \label{PDF_g}
f_{gA} (x)  = \mathbbm{1}(x\geq 0) \frac{1}{g \upsilon} \text{exp} \left(-\frac{x}{g \upsilon}\right)
\end{equation}

The cumulative density function (CDF) of the summation of the two independent RVs of $g_1 A_1$ and $g_2 A_2$, i.e., $g_1 A_1 + g_2 A_2$, for the value $\frac{\gamma_e }{ \rho^2 }$ is
\ifdefined\TwoCol
{\scriptsize
\begin{equation} \label{CDF_Z}
F_Z\left(\frac{\gamma_e }{ \rho^2 }\right)= \frac{1}{g_1 \upsilon_1}  \frac{1}{g_2 \upsilon_2}
\int_{0}^{\frac{\gamma_e }{ \rho^2 }} \int_{0}^{\frac{\gamma_e }{ \rho^2 }-y} \text{exp} \left(-\frac{x}{g_1 \upsilon_1}\right)  \text{exp} \left(-\frac{y}{g_2 \upsilon_2}\right) dx dy ,
\end{equation}
}
\fi
\ifdefined\OneCol 
\begin{align*} 
&F_Z\left(\frac{\gamma_e }{ \rho^2 }\right)=\\
&\frac{1}{g_1 \upsilon_1}  \frac{1}{g_2 \upsilon_2}
\int_{0}^{\frac{\gamma_e }{ \rho^2 }} \int_{0}^{\frac{\gamma_e }{ \rho^2 }-y} \text{exp} \left(-\frac{x}{g_1 \upsilon_1}\right)  \text{exp} \left(-\frac{y}{g_2 \upsilon_2}\right) dx dy , \numberthis \label{CDF_Z}
\end{align*}
\fi
knowing that $ x,y \geq 0$. Solving the double integral, we have
\ifdefined\TwoCol 
{\scriptsize
\begin{equation}
F_Z\left(\frac{\gamma_e }{ \rho^2 }\right)=\frac{g_2 \upsilon_2 \left(\text{exp}\left(-\frac{\gamma_e}{\rho^2 g_2 \upsilon_2}\right)-1\right)+g_1 \upsilon_1 \left(1-\text{exp}\left(-\frac{\gamma_e}{\rho^2 g_1 \upsilon_1}\right)\right)}{g_1\upsilon_1 - g_2 \upsilon_2}.
\end{equation}
}
\fi
\ifdefined\OneCol 
\begin{align*}
&F_Z\left(\frac{\gamma_e }{ \rho^2 }\right)=\\
&\frac{g_2 \upsilon_2 \left(\text{exp}\left(-\frac{\gamma_e}{\rho^2 g_2 \upsilon_2}\right)-1\right)+g_1 \upsilon_1 \left(1-\text{exp}\left(-\frac{\gamma_e}{\rho^2 g_1 \upsilon_1}\right)\right)}{g_1\upsilon_1 - g_2 \upsilon_2}. \numberthis
\end{align*}
\fi
Replacing from (\ref{1minusT_e12_App}) in (\ref{CDF_Z}), we obtain (\ref{T_e12}).

\bibliographystyle{ieeetr}
\bibliography{bibliography}

\end{document}